\documentclass[preprint,12pt]{elsarticle}
\usepackage[top=1in,bottom=1in,left=1in,right=1in]{geometry}
\usepackage{amssymb}
\usepackage{amsmath}
\usepackage{mathtools}
\usepackage{xcolor}
\usepackage{subcaption}
\usepackage{hyperref}
\usepackage{makecell}
\usepackage{booktabs}
\usepackage{upgreek}
\usepackage{tikz}
\usepackage{float} 
\usepackage{natbib}

\usepackage{bm}   

\usepackage{adjustbox} 
\usepackage{multirow} 
\usepackage{cleveref}
\usepackage{amsthm}

\newtheorem{theorem}{Theorem}

\begin{document}

\begin{frontmatter}

\title{A holomorphic Kolmogorov-Arnold network framework for solving elliptic problems on arbitrary 2D domains}

\author[dtuelectro]{Matteo Calafà\texorpdfstring{\fnref{1}}{}}
\address[dtuelectro]{Acoustic Technology, Department of Electrical and Photonics Engineering, Technical University of Denmark, 2800 Kongens Lyngby, Denmark}
\author[au]{Tito Andriollo}
\address[au]{Mechanics and Materials, Department of Mechanical and Production Engineering, Aarhus University, 8200 Aarhus, Denmark}
\author[dtucompute]{Allan P. Engsig-Karup}
\address[dtucompute]{Scientific Computing, Department of Applied Mathematics and Computer Science, Technical University of Denmark, 2800 Kongens Lyngby, Denmark}
\author[dtuelectro]{Cheol-Ho Jeong}

\fntext[1]{Formerly at: Mechanics and Materials, Department of Mechanical and Production Engineering, Aarhus University, 8200 Aarhus, Denmark \\ Corresponding author: matcal@dtu.dk}

\begin{abstract}
Physics-informed holomorphic neural networks (PIHNNs) have recently emerged as efficient surrogate models for solving differential problems. By embedding the underlying problem structure into the network, PIHNNs require training only to satisfy boundary conditions, often resulting in significantly improved accuracy and computational efficiency compared to traditional physics-informed neural networks (PINNs). In this work, we improve and extend the application of PIHNNs to two-dimensional problems. First, we introduce a novel holomorphic network architecture based on the Kolmogorov-Arnold representation (PIHKAN), which achieves higher accuracy with reduced model complexity. Second, we develop mathematical extensions that broaden the applicability of PIHNNs to a wider class of elliptic partial differential equations, including the Helmholtz equation. Finally, we propose a new method based on Laurent series theory that enables the application of holomorphic networks to multiply-connected plane domains, thereby removing the previous limitation to simply-connected geometries.
\end{abstract}


\end{frontmatter}

\section{Introduction}
Multi-layer perceptrons (MLPs) are a simple yet fundamental class of neural networks (NNs) that have played a significant role in advancing modern deep learning \cite{lecun2015deep, kruse2011computational}. MLPs are fully-connected feed-forward neural networks with non-linear activation functions, and their importance is rooted in the universal approximation theorem 
\cite{cybenko1989approximation}, which guarantees the ability to approximate any continuous function with error bounds. 

MLPs have become a standard choice in physics-informed machine learning \cite{karniadakis2021physics}, where neural network training is guided by physical laws to resolve key challenges such as ensuring physical consistency, generalization beyond training data and addressing limited data availability. Specifically, MLPs have been employed to define physics-informed neural networks (PINNs), starting from early studies \cite{Lee1990NeuralEquations,PsichogiosEtAl1992,
Dissanayake1994, Meade1994SolutionNetworks, Meade1994TheNetworks, lagaris1998artificial} and continuing through their recent surge in popularity \cite{raissi2019physics,cuomo2022scientific}. 

In contrast to this trend, Kolmogorov-Arnold networks (KANs) have recently been proposed as an alternative to classical MLPs \cite{liu2024kan}, inspired by the Kolmogorov-Arnold representation theorem \cite{kolmogorov1957representation}. Specifically, KANs consist of multiple stacked layers, with each layer corresponding to the formula in the representation theorem. In this framework, activation functions are adaptive, in contrast with MLPs, where activation functions are fixed and only the linear weights are trained. KANs have proven to overcome MLPs in different scenarios, especially due to the decreased network size \cite{vaca2024kolmogorov,elaziz2024}, higher interpretability \cite{liu2024kan2,xu2024kolmogorov,zhang2025physics} and robustness \cite{dong2024kolmogorov}.

On the other hand, the B-spline activation functions require the evaluation on a grid \cite{unser1993b}, which inevitably increases the computational time and resources. For this reason, subsequent works proposed different activation functions, such as wavelets \cite{bozorgasl2024wav}, radial basis functions \cite{li2024kolmogorov} and Chebyshev polynomials \cite{ss2024chebyshev}.

KANs have also been adopted to learn from differential equations. Specifically, a first test with physics-informed KANs (PIKANs) is performed in the original paper \cite{liu2024kan}, whereas \citet{abueidda2025deepokan} employed KANs as deep operators (DeepONets \cite{lu2021learning}). A comprehensive comparison between PIKANs and traditional PINNs was presented by \citet{toscano2025pinns} and \citet{shukla2024}, showing that PIKANs can achieve comparable or better accuracy when Chebyshev polynomials are employed, along with an additional stabilization mechanism. Additionally, \citet{koenig2024kan} proposed a KAN architecture for learning solutions to dynamical systems.

Recently, \citet{calafa2024} systematically introduced physics-informed holomorphic neural networks (PIHNNs), as a powerful approach to accelerating training for problems that can be expressed using complex holomorphic potentials. By enforcing holomorphic outputs, PIHNNs can automatically satisfy the governing equations for models such as the two-dimensional Laplace problem and linear elasticity. Consequently, the network training is performed only to fulfill the boundary conditions (BCs), leading to significant improvements in both accuracy and efficiency. A modified version of PIHNNs was also proposed \cite{calafa2025solving} to address plane crack problems in linear elasticity by incorporating singularities at the cracks.

On the other hand, vanilla holomorphic networks present three main limitations: first, they are defined as complex-valued MLPs, with each layer performing a holomorphic operation. Therefore, the underlying architecture necessitates the use of analytic activation functions (which are unbounded by Liouville's theorem) such as the exponential. This implies careful consideration about the stability of the network, including ad-hoc weight initializations and moderate learning rates. Additionally, the method applies only to simply-connected domains, requiring the employment of domain-decomposition strategies for general multiply-connected domains, which can make the implementation and optimization problem more challenging. Finally, the original method has been applied only to a few classes of problems, i.e., the 2D Laplace problem and linear elasticity.

\subsection{Contributions}
In this work, we propose several techniques to address the limitations outlined previously. First, we introduce a novel KAN-based architecture which seamlessly integrates into the holomorphic framework and avoid to employ rapidly diverging exponential activation functions. The new architecture, referred to as physics-informed holomorphic Kolmogorov-Arnold network (PIHKAN), is selected due to its spectral properties, and a suitable weight initialization is proposed. Then, we illustrate how to extend the holomorphic network approach to a broader class of two-dimensional elliptic problems through the use of Vekua operators, including the Helmholtz problem. In addition, we propose a new method based on the Laurent theorem to effectively handle multiply-connected domains, which eliminates the need for domain decomposition and achieves rapid convergence.

\subsection{Paper organization}
The article is organized as follows: in \Cref{sec:methods}, we present the mathematical formulation of the target differential problems and describe the holomorphic neural network framework to solve these. Specifically, we introduce the architectures of both MLPs and KANs, along with their holomorphic adaptations. We also detail the proposed PIHKAN weight initialization method and the Laurent-based approach for multiply-connected domains. \Cref{sec:numericaltests} presents three numerical experiments, where the performance of PIHNNs and PIHKANs is evaluated and compared with that of classical PINNs, as well as with each other. Finally, in \Cref{sec:conclusions} we summarize the outcomes of this research and elucidate future goals.

\section{Methods}\label{sec:methods}
This section describes the proposed learning strategy. \Cref{sec:holomorphic representation} outlines the problems we address and the mathematical properties we leverage, \Cref{sec:physics-informed holomorphic machine learning} introduces the main concepts of holomorphic neural networks while \Cref{sec:kolmogorov-arnold networks} presents the proposed network architecture through KANs and a suitable weight initialization strategy. Finally, we present in \Cref{sec:laurent} the Laurent-based method for handling multiply-connected domains.

\subsection{Holomorphic representation of PDE solutions}\label{sec:holomorphic representation}
Let us consider a two-dimensional compact domain $\Omega \subset \mathbb{R}^2$, and an unknown smooth function $\bm{u}:\Omega \rightarrow \mathbb{R}^{N_u}$  of dimension $N_u \in \mathbb{N}$. We study problems in the following form:
\begin{equation}\label{eq:general_problem}
    \begin{cases}
        \mathcal{F}[\bm{u}](\bm{x}) = \bm{0}, & \bm{x} \in \Omega, \\
        \mathcal{B}[\bm{u}](\bm{x}) = \bm{0}, & \bm{x} \in \partial \Omega,
    \end{cases}
\end{equation}
where $\mathcal{F}$ describes a set of partial differential operators while $\mathcal{B}$ embodies the applied boundary conditions. 

Let us denote with $H(\Omega)$ the set of holomorphic functions on $\Omega$, i.e., $\varphi \in H(\Omega)$ if and only if

\begin{equation*}
    \lim_{z\rightarrow z_0} \frac{\varphi(z)-\varphi(z_0)}{z-z_0},
\end{equation*}
exists for every $z\in\mathbb{C}$ such that $(\text{Re}(z),\text{Im}(z))\in\Omega$. A problem in the form of \Cref{eq:general_problem} is said to admit a holomorphic representation if there exists an operator $\mathcal{F}^*$ and $N_*\in\mathbb{N}$ such that the kernel of $\mathcal{F}$ is equal to the image of $\mathcal{F}^*$ through $[H(\Omega)]^{N_*}$. Namely, the formulation in \Cref{eq:general_problem} can be replaced by the following: find $\varphi_1,\dots,\varphi_{N_*} \in H(\Omega)$ such that

\begin{equation}\label{eq:holo_problem}
        \mathcal{B}\left[\mathcal{F}^*[\varphi_1,\cdots,\varphi_{N_*}]\right](\bm{x}) = \bm{0}, \; \;\forall \bm{x} \in \partial \Omega.
\end{equation}

\Cref{tab:holo_problems} presents some 2D problems that admit a holomorphic representation. Each case arises from the fundamental connection between holomorphic and harmonic functions and is valid only when $\Omega$ is simply connected.

\begin{table}[!ht]
\small
    \centering
    \renewcommand{\arraystretch}{1.2}
    \begin{tabular}{c c c c c c}
        \toprule
        \makecell{Problem \\ name} & $N_u$ & $N_*$ & $\mathcal{F}[\bm{u}]$ & $\mathcal{F}^*[\varphi_1,\dots,\varphi_{N_*}]$  & Application \\
        \midrule \midrule
         \makecell{Laplace \\ equation} & 1 & 1 & $\nabla^2 \bm{u}$ & $\text{Re}(\varphi_1)$ & \makecell{Potential theory, \\ equilibrium}\\ 
         \midrule
         \makecell{Biharmonic \\ equation} & 1 & 2 & $\nabla^4\bm{u}$ & $\text{Re}(\overline{z} \varphi_1 + \varphi_2)$ & \makecell{Plate bending, \\ elasticity}\\
         \midrule
         \makecell{Linear \\ elasticity} & 2 & 2 & \makecell{$\mu \nabla^2\bm{u} +$ \\$\left(\tilde{\lambda} + \mu\right) \nabla(\nabla \cdot \bm{u})$} & $ \begin{bmatrix}
          \frac{1}{2\mu}\text{Re}\left(\kappa\varphi_1 - z \overline{\varphi_1'} - \overline{\varphi_2}\right)\\
        \frac{1}{2\mu}\text{Im}\left(\kappa\varphi_1 - z \overline{\varphi_1'} - \overline{\varphi_2}\right)\end{bmatrix}$ & \makecell{Stress analysis, \\ deformation} \\
        \midrule
        \makecell{Helmholtz \\ equation} & 1 & 1 & $\nabla^2 \bm{u} + \beta^2 \bm{u} $ & $\mathcal{V}_\beta[\text{Re}(\varphi_1)]$ & \makecell{Wave propagation, \\ vibrations} \\
        \bottomrule
    \end{tabular}
    \caption{Some 2D differential problems satisfying the holomorphic potential representation.}
    \label{tab:holo_problems}
\end{table}

The Laplace equation provides a simple yet fundamental model to describe systems in equilibrium state, such as in electrostatics. Its solutions  are called harmonic functions, and their holomorphic representation is a basic result from complex analysis \cite{rudin1987real}. This concept extends to the biharmonic equation due to Goursat's theorem \cite{goursat1898}, as shown in the second row of \Cref{tab:holo_problems}. This model is often adopted to analyze the bending of thin solid plates \cite{selvadurai2013partial}. 

Linear elasticity offers a more general framework for modeling the internal stress and deformation of solid materials, and it is one of the most important models in solid mechanics \cite{gould1994introduction}. The holomorphic representation of linear elasticity arises from the biharmonic nature of the Airy stress function, as developed by Kolosov \cite{kolosov1909application}. In this formulation, $\lambda,\mu$ are the first and second Lamé parameters, $\tilde{\lambda} = \lambda$ for plane strain, whereas $\tilde{\lambda} = (2\lambda\mu) / (\lambda + 2\mu)$ for plane stress. Furthermore, $\kappa = (\tilde{\lambda} + 3\mu)/(\tilde{\lambda} + \mu)$ is the Kolosov constant. Additionally, the holomorphic representation of stresses and other secondary variables can also be readily derived \cite{muskhelishvili1977}. 

This study further extends the holomorphic approach to the Helmholtz equation, a fundamental model to describe wave propagation phenomena in fields like acoustics, optics,  quantum mechanics, and water wave mechanics  \cite{bayliss1983iterative,wang1997helmholtz,haber2011fast,DeanDalrymple1991}. To the best of our knowledge, this constitutes the first application of holomorphic neural network to such problem. It is worth noting that recent studies \cite{qu2024boundary,schmid2025boundary} have investigated PINNs with boundary formulations for acoustic applications, showing promising results. However, the problems considered in these works are often relatively simple, particularly with respect to low wave numbers. The holomorphic representation of the Helmholtz equation is given in the final row of \Cref{tab:holo_problems}, where $\beta$ is any positive constant representing the wave number and $\mathcal{V}_\beta$ is the operator defined as

\begin{equation}\label{eq:vekua}
    \mathcal{V}_\beta[g](\bm{x}):= g(\bm{x}) - \int_{0}^1 \frac{J_1(\beta|\bm{x}|\sqrt{1-t})}{2\sqrt{1-t}}  \beta|\bm{x}| \, g(t\bm{x}) \, dt,
\end{equation}

where $J_1$ is the first order Bessel function of the first kind. Specifically, $\mathcal{V}_\beta$ maps harmonic functions to solutions of the Helmholtz equation and the associated holomorphic representation is valid under minimal hypotheses, as shown in \citet{moiola2011vekua}. More broadly, $\mathcal{V}_\beta$ belongs to the class of Vekua operators, introduced by \citet{vekua1967}, which serve to map harmonic functions to solutions of more general elliptic partial differential equations. However, explicit forms of such operators are currently known for only a few specific cases, including the Helmholtz equation, and extending this framework to other problems remains an active area of research \cite{henrici1957survey}. Consequently, the list in \Cref{tab:holo_problems} may be expandable to a wider class of elliptic problems.

Further, due to the linearity of the PDEs in \Cref{tab:holo_problems}, the holomorphic representation can be extended to non-homogeneous problems. If $\mathcal{F}[\bm{u}] = \bm{b}$ for $\bm{b}\neq \bm{0}$, then 
\begin{equation}\label{eq:non-homo}
    \bm{u} = \mathcal{F}^*[\varphi_1,\dots,\varphi_{N_*}] + \mathcal{F}^{-1}[\bm{b}],
\end{equation}
is the set of all solutions, where $\mathcal{F}^{-1}[\bm{b}]$ identifies a particular solution to the problem that does not necessarily satisfy the boundary conditions. Specifically, $\mathcal{F}^{-1}[\bm{b}]$ could be obtained analytically, numerically or through the Green's function method.

\subsection{Physics-informed holomorphic machine learning}\label{sec:physics-informed holomorphic machine learning}
In the following, we denote by $\mathbb{S}$ either the field of real numbers $\mathbb{R}$ or the field of complex numbers $\mathbb{C}$. PINNs are usually constructed as fully-connected feed-forward MLP networks. Specifically, a vanilla $\mathbb{S}$-valued MLP of $L\in\mathbb{N}$ layers and $\{M_l\}_{l=0}^{L} \subset \mathbb{N}$ neurons per layer is defined as a composite function 
\begin{equation}\label{eq:MLP}
    \text{MLP}(\bm{x}):=\mathcal{T}_L \circ \phi \circ \mathcal{T}_{L-1} \circ \dots \circ \phi \circ \mathcal{T}_1(\bm{x}),
\end{equation}
where $\phi:\mathbb{S}\rightarrow \mathbb{S}$ is the (non-trainable) activation function and $\mathcal{T}_{l}$ is the linear operator $\mathcal{T}_l:\mathbb{S}^{M_{l-1}} \rightarrow \mathbb{S}^{M_{l}}$, $\mathcal{T}_l(\bm{x}) = \bm{W}_l \bm{x} + \bm{b}_l$, identified by the trainable weight matrix $\bm{W}_l \in \mathbb{S}^{M_{l},M_{l-1}}$ and bias vector $\bm{b}_l \in \mathbb{S}^{M_l}$. In particular, $M_0$ is the input dimension and $M_L$ is the output dimension. In the real-valued case, common choices of $\phi$ are the hyperbolic tangent \cite{anastassiou2011univariate}, the rectified linear unit ReLU$(x):=\max\{0,x\}$ \cite{li2017convergence} and the sigmoid function \cite{han1995influence}. For complex-valued networks (CVNNs), split activation functions 
\begin{equation}\label{eq:split_activation}
    \phi(z):=\phi_\mathbb{R}\left(x\right) + i \phi_\mathbb{R}\left(y\right),
\end{equation}
 are often adopted \cite{nitta1997extension,hirose2012complex,trabelsi2017deep}, where $\phi_\mathbb{R}$ is a real-valued activation function such as those mentioned above and $x,y$ are the real and imaginary part of $z$, respectively.
 
PINNs can readily embed physical models from \Cref{eq:general_problem} into a machine learning framework \cite{raissi2019physics}. Specifically, one can define the loss function $\mathcal{L}$ as 
\begin{equation}\label{eq:loss_pinn}
    \mathcal{L}[\bm{u}] =  \mathbb{E}_{\bm{x} \sim \mathcal{P}_\Omega} \left[\bm{\lambda}_\mathcal{F} \cdot |\mathcal{F}[\bm{u}]|^2\right] +  \mathbb{E}_{\bm{x} \sim \mathcal{P}_{\partial\Omega}} \left[\bm{\lambda}_\mathcal{B} \cdot |\mathcal{B}[\bm{u}]|^2\right],
\end{equation}
where $\bm{\lambda}_\mathcal{F},\bm{\lambda}_\mathcal{B} >0$ are hyperparameters, and $\mathcal{P}_{\Omega},\mathcal{P}_{\partial\Omega}$ denote distributions to sample training points from $\Omega$ and $\partial\Omega$, respectively. Thus, the gradient of $\mathcal{L}$ with respect to all network weights can be efficiently obtained via backpropagation. Optimizers like Adam \cite{kingma2014adam} or LBFGS \cite{liu1989limited} then use this information to update the weights and minimize the loss.

If $\mathbb{S}\equiv \mathbb{C}$, the complex derivative is typically computed by software libraries through the Wirtinger derivative \cite{wirtinger1927formalen}
\begin{equation}\label{eq:wirtinger}
    \frac{\partial \mathcal{L}}{\partial z} = \frac{1}{2}\left(\frac{\partial \mathcal{L}}{\partial x} - i \frac{\partial \mathcal{L}}{\partial y}\right).
\end{equation}
This choice arises from the need of performing the backpropagation even when the network is not strictly holomorphic, such as when the split activation in \Cref{eq:split_activation} is employed \cite{trabelsi2017deep}. Indeed, \Cref{eq:wirtinger} only requires differentiability with respect to $x,y$ separately, without taking into account the validity of the Cauchy-Riemann equations. Furthermore, the definition is consistent since the two derivatives coincide when the function is holomorphic.

Following \citet{calafa2024}, we call physics-informed holomorphic neural networks (PIHNNs) the $\mathbb{C}$-valued MLPs from \Cref{eq:MLP} such that $\phi \in H(\mathbb{C})$. PIHNNs generate only $H(\mathbb{C})$ functions and can be proven to satisfy the universal approximation theorem under the compact uniform convergence \cite{park2024qualitative}. PIHNNs can thus be adopted to learn from the problem in \Cref{eq:holo_problem}. Specifically, the loss can be suitably defined as

\begin{equation}\label{eq:loss_pihnn}
    \mathcal{L}[\varphi_1,\dots,\varphi_{N_*}] =  \mathbb{E}_{\bm{x} \sim \mathcal{P}_{\partial\Omega}} \left[\bm{\lambda}_\mathcal{B} \cdot |\mathcal{B}[\mathcal{F}^*[\varphi_1,\dots,\varphi_{N^*}]]|^2\right],
\end{equation}
where $\varphi_1,\dots,\varphi_{N^*}$ are now the outputs from $N^*$ independent PIHNNs.

By comparing \Cref{eq:loss_pinn} with \Cref{eq:loss_pihnn}, one can readily notice the several advantages offered by the holomorphic representation: first, \Cref{eq:holo_problem} imposes a hard constraint on the PDE(s), improving the accuracy of the learned solution. Additionally, training points are placed exclusively on the boundary $\partial \Omega$, enhancing the efficiency and memory requirements. Finally, the number of hyperparameters $\bm{\lambda}$ is reduced, simplifying the fine-tuning process.

To the best of the authors' knowledge, PIHNNs have been adopted only in a few previous studies \cite{ghosh2023harmonic,calafa2024,calafa2025solving}. The first of these focuses solely on the Laplace problem in \Cref{tab:holo_problems}, while the latter two are primarily addressed to linear elasticity. Additionally, these latter works provide a more comprehensive analysis, including a proof of the universal approximation theorem and a suitable weight initialization strategy. In terms of activation functions, the first study employs $\phi(z)=\sin(z)$, whereas the latter two utilize $\phi(z)=e^z$, demonstrating its preferable properties.

\subsection{Kolmogorov-Arnold networks}\label{sec:kolmogorov-arnold networks}

The recently proposed KANs \cite{liu2024kan} are inspired by the Kolmogorov-Arnold representation theorem \cite{kolmogorov1957representation}, which states that any multi-valued continuous function $f:[0,1]^M \rightarrow \mathbb{R}$ can be written in terms of some scalar functions $f_j: \mathbb{R} \rightarrow \mathbb{R}$ and $f_{j,k}: [0,1] \rightarrow \mathbb{R}$ as

 \begin{equation}\label{eq:Kolmogorov-Arnold-theorem}
     f(x_1,x_2,\dots,x_M) = \sum_{j=0}^{2M} f_j \left( \sum_{k=1}^M f_{j,k}(x_k)\right).
 \end{equation}

\begin{figure}[!ht]
    \centering
    \vspace{5mm}
    \includegraphics[width=1\linewidth]{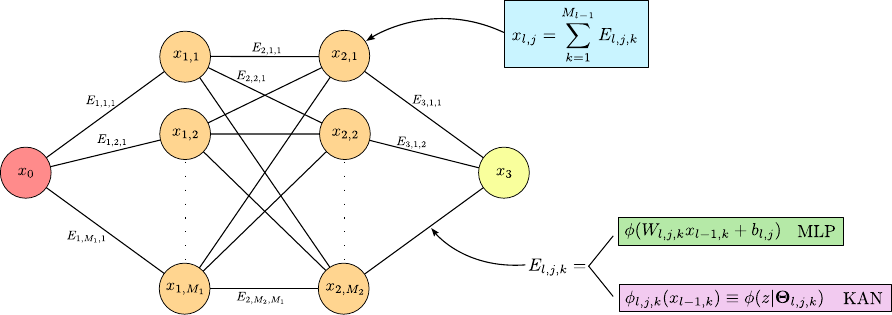}
    \vspace{4mm}
    \caption{Visualization of multi-layer feed-forward network. $E_{l,j,k}$ denotes the edge connecting the $k$-th neuron at the $(l-1)$-th layer to the $j$-th neuron at the $l$-th layer. Each node computes the sum of the evaluations from each input edge. In MLPs, the evaluation on $E_{l,j,k}$ is $\phi(W_{l,j,k}x_{l-1,k} + b_{l,j})$, with $W_{l,j,k}, b_{l,j}$ being trainable scalar parameters. In KANs, the evaluation is instead $\phi_{l,j,k}(x_{l-1,k}$), with $\phi_{l,j,k}$ being a trainable scalar activation function that is parametrized by a set of weights ${\bf\Theta}_{l,j,k}$.}
    \label{fig:nn}
\end{figure}

 The idea proposed in \citet{liu2024kan} is to employ \Cref{eq:Kolmogorov-Arnold-theorem} while stacking more layers. Namely, a KAN network of $L\in\mathbb{N}$ layers is defined as

 \begin{equation}\label{eq:KAN}
     \text{KAN}(\bm{x}):=\phi_L \circ \phi_{L-1} \circ \dots \circ \phi_1(\bm{x}),
 \end{equation}

 where each layer $l=1,\dots,L$ performs the operation
 \begin{equation*}
     x_{l,j} = \sum_{k=1}^{M_{l-1}}\phi_{l,j,k}(x_{l-1,k}), \hspace{1cm} j=1,\dots,M_{l+1},
 \end{equation*}
being $x_{l,j} \in \mathbb{S}$ the $j$-th network activation at the $l$-th layer and $\phi_{l,j,k}:\mathbb{S}\rightarrow\mathbb{S}$ a trainable activation function. To enable learning, $\phi_{l,j,k}$ is a function parameterized by a set of trainable weights ${\bf \Theta}_{l,j,k}$. Comparing \Cref{eq:MLP} with \Cref{eq:Kolmogorov-Arnold-theorem}, the key difference between MLPs and KANs is that KANs lack linear operators $\mathcal{T}_l$, with trainable parameters instead embedded in the learnable activation functions (see \Cref{fig:nn}).

As already mentioned, \citet{liu2024kan} originally proposed to construct the learnable activation functions as B-spline curves. This implies that $G+P$ scalar parameters are employed for each activation $\phi_{l,j,k}$, where $G\in\mathbb{N}$ is the number of B-splines grid intervals and $P\in\mathbb{N}$ is the maximum polynomial degree. Conversely, the representation in terms of Chebyshev polynomials only employs $P$ parameters per activation function \cite{ss2024chebyshev,shukla2024}.

Apart from the architectural differences, the learning algorithm remains the same for both MLPs and KANs. Physics-informed KANs (PIKANs) are defined by \Cref{eq:KAN} with a loss function given by \Cref{eq:loss_pinn}. Furthermore, we call PIHKANs the $\mathbb{C}$-valued KANs in \Cref{eq:KAN}, whose trainable activations $\phi_{l,j,k}$ are all holomorphic and whose loss is given by \Cref{eq:loss_pihnn}. 

It is note of worth that PIHKANs cannot be directly constructed as $\mathbb{C}$-valued vanilla KANs, as B-splines are only continuously differentiable up to order $P$ and are therefore not analytic. This limitation is analogous to the case of PIHNNs, which are not defined as standard $\mathbb{C}$-valued PINNs due to the fact that common activation functions (e.g., sigmoid, $\tanh$, ReLU) are not holomorphic. On the other hand, when polynomials are used as activation functions, real-valued networks can be naturally extended to form holomorphic networks. While polynomial activation functions are generally unsuitable for traditional PINNs, they have been found to offer significant benefits in the context of KANs, where Chebyshev polynomials have recently demonstrated notable advantages \cite{ss2024chebyshev,shukla2024}.

Motivated by these findings, we select a family of polynomials that satisfies certain $L^2$ orthogonality properties on the complex plane, rather than the real line. Specifically, we select the monomial basis $\mathcal{M}_P = \{1,z,\dots,z^{P}\}$ based on the following facts: 1) It generates all complex polynomials until order $P$. 2) It is $L^2$-orthogonal on the unit disc $\mathcal{D}=\{z \in \mathbb{C}: |z|\le 1\}$. 3) It is $L^2$-orthogonal on the unit circumference $\partial \mathcal{D}$. 4) It provides high interpretability of the representation. 5) It is mathematically and computationally efficient to handle. Thus, we define the PIHKAN layer as

\begin{equation}\label{eq:monomials}
    \phi_{l,j,k}(z) \equiv \phi(z|{\bf \Theta}_{l,j,k}) := \sum_{p=0}^{P}W_{l,j,k,p} z^p = \sum_{p=1}^{P}W_{l,j,k,p} z^p + b_{l,j,k}, 
\end{equation}

where ${\bf \Theta}_{l,j,k} = \{W_{l,j,k,p},b_{l,j,k}\}_{p=1}^P$ are the trainable complex-valued weights and biases.
    
We further propose a suitable weight initialization strategy to ensure stability during both the forward and backward passes. Proper weight initialization is crucial for improving training performance and mitigating the risk of exploding or vanishing gradients \cite{narkhede2022review}, particularly in this context where activation functions can rapidly diverge. We first emphasize that previous works on KANs have mostly employed empirical initializations rather than theoretically rigorous. For example, \citet{liu2024kan} employs normal distributions with a fixed small variance ($\sigma =0.1$), meanwhile different other works \cite{abueidda2025deepokan,ss2024chebyshev,shukla2024} improperly adopt the Xavier normal initialization \cite{glorot2010understanding}.

Similarly to \citet{glorot2010understanding} and \citet{he2015delving}, we adopt a criterion for weight initialization based on maintaining variance stability across layers. Specifically, all bias vectors $b_{l,j,k}$ are initialized to zero, consistent with standard practices. Then, let us assume that the evaluation of the PIHKAN at the $(l-1)$-th layer is distributed according to $z_{l-1,k} \sim \mathcal{N}(0,1)$ independently for each neuron $k=1,\dots,M_{l}$ and that each weight is sampled according to $W_{l,j,k,p} \sim \mathcal{N}(0,\sigma_{p}^2)$ independently for each $j,k,p$. Here, $\{\sigma_p\}_{p=1}^{P}>0$ needs to be tuned by imposing that $\mathbb{V}[z_{l,j}] = 1$ for each $j$, where we denote with $\mathbb{V}[z_{l,j}]$ the variance of the random variable $z_{l,j}$. If this condition is satisfied, recursion and the central limit theorem guarantee that the variance remains stable, approximately unitary across the layers of the PIHKAN. Namely, we have for each neuron $j=1,\dots,M_{l+1}$:
\begin{equation*}
\begin{aligned}
    \mathbb{V}\left[z_{l,j}\right] = \mathbb{V}\left[\sum_{k=1}^{M_{l-1}}\phi_{l,j,k}(z_{l-1,k})\right] = \mathbb{V}\left[\sum_{k=1}^{M_{l-1}} \sum_{p=1}^{P}W_{l,j,k,p}z_{l-1,k}^p\right] = \sum_{k=1}^{M_{l-1}} \sum_{p=1}^{P} \sigma^2_p \cdot p! .
\end{aligned}
\end{equation*}

In the previous steps, we leveraged the independence assumptions and the property that $\mathbb{V}[z^p] = p!$ when $z \sim \mathcal{N}(0,1)$. It is curious to note that this result differs and becomes less straightforward when the input is real-valued.  

The variance at the $l$-th layer is unitary if we select $\sigma^2_p = 1/( M_{l-1}\cdot  P\cdot p! )$. This choice aligns with the Xavier initialization \cite{glorot2010understanding}, except for the additional factor $P\cdot p!$.

To guarantee stability in the backward pass as well, we can apply similar passages and obtain

\begin{equation*}
\begin{aligned}
    \mathbb{V}\left[\frac{\partial \mathcal{L}}{\partial z_{l-1,k}}\right] = \mathbb{V}\left[\sum_{j=1}^{M_{l}}\frac{\partial \phi_{l,j,k}(z_{l-1,k})} {\partial z_{l-1,k}} \frac{\partial \mathcal{L}}{\partial z_{l,j}}\right] = \mathbb{V}\left[\sum_{j=1}^{M_{l}} \sum_{p=1}^{P}W_{l,j,k,p}pz_{l-1,k}^{p-1}\right] = \sum_{k=1}^{M_{l}} \sum_{p=1}^{P} \sigma^2_p \cdot p \cdot p!,
\end{aligned}
\end{equation*}

where $\mathcal{L}$ denotes the loss function and the variance of $\partial \mathcal{L}/\partial z_{l,j}$ is assumed to be unitary. The preceeding analysis indicates that the choice $\sigma^2_p = 1/( M_{l} \cdot P\cdot p\cdot p! )$ ensures stability in the backward pass. However, this is generally incompatible with the prescribed value for the forward pass. To address this discrepancy, we adopt the heuristic proposed by \citet{glorot2010understanding}, which applies an appropriate average between the two. Therefore, we get

\begin{equation}\label{eq:initialization}  
    \text{Re}(W_{l,j,k,p}),\text{Im}(W_{l,j,k,p}) \overset{\mathrm{iid}}{\sim} \mathcal{N}\left(0, \frac{2}{(M_{l-1} + pM_{l})\cdot p!\cdot P}\right).  
\end{equation}  

The weight initialization strategy is effective if the input is approximately $z_0 \sim \mathcal{N}(0,1)$. Input normalization \cite{sola1997importance} can be in principle be employed to ensure the zero mean and unitary variance but  the operation is not holomorphic. To overcome this issue, we perform the following trick: first, we select a sampling distribution $\mathcal{P}_{\partial \Omega}$ (see \Cref{eq:loss_pihnn}). Then, a large number of points is sampled according to $\mathcal{P}_{\partial \Omega}$ (indicatively, 10 times the cardinality of the training set) and is used to estimate the average $\mu_{\mathcal{P}_{\partial\Omega}}$ and the variance $\sigma^2_{\mathcal{P}_{\partial\Omega}}$. These parameters are fixed, and the normalization
\begin{equation}\label{eq:normalization}
    \tilde{z} = \frac{z-\mu_{\mathcal{P}_{\partial\Omega}}}{\sigma_{\mathcal{P}_{\partial\Omega}}},
\end{equation}
is applied at the first layer of the network, for any input $z \in \mathbb{C}$. Despite training points need to be initially sampled according to $P_{\partial \Omega}$ to comply with the weight initialization,
adaptive refinement or re-sampling can be adopted at a later stage.

\subsection{Laurent-based extension to multiply-connected domains}\label{sec:laurent}

The proposed strategy is effective only if $\Omega$ is simply-connected, meaning in practice that it cannot contain internal holes. This represents a crucial limitation of both PIHNNs and PIHKANs since many problems of interest are set on multiply-connected domains. For instance, linear elasticity is often applied in fracture mechanics, where small holes naturally originate within the solid domain \cite{anderson2005,calafa2024}. Previous studies on PIHNNs \cite{ghosh2023harmonic,calafa2024,calafa2025solving} employed domain decomposition to overcome the limitation in a framework equivalent to extended PINNs (XPINNs) \cite{jagtap2020extendedphysicsinformed}. However, the addition of domain interfaces and multiple parallel networks inevitably increase the computational cost and complicate the convergence process. In this section, we illustrate an alternative and efficient strategy to deal with general domains.

First, it is important to emphasize that standard PIHNNs and PIHKANs are limited to simply-connected domains for two distinct reasons. The first stems from the assumptions required for the holomorphic representations listed in \Cref{tab:holo_problems}. The second relates to the conditions necessary to guarantee the universal approximation property of holomorphic neural networks. These two issues must therefore be addressed separately. The first can be resolved directly by the following theorem, originally stated by \citet{axler1986harmonic}.

\begin{figure}[!ht]
    \centering
    \includegraphics[width=0.5\linewidth]{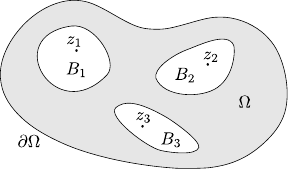}
    \caption{Sketch of multiply-connected domain with $S=3$ holes.}
    \label{fig:multiply_connected}
\end{figure}

\begin{theorem}[Logarithmic conjugation]\label{theo:logarithmic}
    Let $\Omega \subset \mathbb{R}^2$ a bounded domain, with $\{B_s\}_{s=1}^S\subset \mathbb{R}^2$ denoting its internal $S\in\mathbb{N}$ holes. Let also $z_s \in B_s$ for every $s=1,\dots,S$ and $\nabla^2u = 0$ on $\Omega$ (see \Cref{fig:multiply_connected}). Then, there exists $\varphi \in H(\Omega)$ and $\{c_s\}_{s=1}^S \subset \mathbb{R}$ such that

    \begin{equation*}
        u(x,y) = \textnormal{Re}(\varphi(z)) + \sum_{s=1}^S c_s \log{|z-z_s|}, \hspace{1cm} z=x+iy, \; (x,y)\in\Omega.
    \end{equation*}
\end{theorem}

\Cref{theo:logarithmic} is in essence the extension of the holomorphic representation for the Laplace problem in \Cref{tab:holo_problems} to multiply-connected domains. Formulated differently, we can maintain the same original representation $u=\text{Re}(\varphi)$ as long as 
\begin{equation}\label{eq:holo_log}
    \varphi(z) = \varphi_0(z) +\sum_{s=1}^S c_s \log(z-z_s),
\end{equation}
where $\varphi_0 \in H(\Omega)$ and $\log(\cdot)$ is now the complex logarithm. Since all representations in \Cref{tab:holo_problems} stem from the holomorphic characterization of harmonic functions, we conclude that applying the transformation in \Cref{eq:holo_log} to each complex potential is sufficient to extend these representations to multiply connected domains. Furthermore, $\{c_s\}_{s=1}^S$ in \Cref{eq:holo_log} can be seamlessly integrated into a machine learning framework as additional trainable parameters.

As mentioned above, standard holomorphic networks are still unable to approximate all holomorphic functions on a multiply connected domain. This limitation arises because holomorphic functions cannot always be represented by Taylor series but require Laurent series instead \cite{howie2003}. Specifically, while the Taylor expansion in the form of $\sum_{j=0}^\infty a_j (z-z_0)^j$ can accurately represent $\varphi_0$ within a disk centered at $z_0 \in \mathbb{C}$, only the Laurent series $\sum_{j=-\infty}^\infty a_j (z-z_0)^j$ can fully capture $\varphi_0$ in an annular region. To integrate the necessary negative powers into the network representation, while complying with \Cref{eq:holo_log}, we introduce the following formulation:

\begin{equation}\label{eq:laurent}
    \varphi_n(z) = \varphi_{n,0}(z) + \sum_{s=1}^S \varphi_{n,s}\left(\frac{1}{z-z_s}\right) + \sum_{s=1}^S c_{n,s} \log (z-z_s), \hspace{1cm} n=1,\dots,N_*.
\end{equation}
Here, $\varphi_n$ represents the $n$-th complex potential that is employed for the problems in \Cref{tab:holo_problems}, $\{\varphi_{n,s}\}_{s=0}^S$ are the outputs of $S+1$ independent holomorphic networks (either PIHNNs or PIHKANs), while $\{c_{n,s}\}_{s=1}^S$ are real-valued scalar trainable parameters.

\Cref{eq:laurent} inherently induces instability for $z\rightarrow z_s$, however $|z-z_s|$ has a lower bound given by the distance between $z_s$ and the border $B_s$. Therefore, $z_s$ should be chosen as the most internal point of $B_s$. In addition, the normalization in \Cref{eq:normalization} can be applied to $\{\varphi_{n,s}\}_{s\ge 1}$ as well, guaranteeing stability at the beginning of the training. It is also important to note that the application of $S+1$ networks for a single domain could seem expensive, but the networks associated to $\{\varphi_{n,s}\}_{s\ge 1}$ are typically smaller than the one associate to the positive powers $\varphi_{n,0}$.

\section{Numerical tests}\label{sec:numericaltests}
We present some experimental tests, where PIHKANs are compared to both PIHNNs and traditional PINNs. The holomorphic KAN architecture has been integrated in the original PIHNN library \footnote{\url{https://github.com/teocala/pihnn}} under \texttt{LGPL 2.1} license, and will be made available after the publication of the article. The backend of the library is \texttt{PyTorch} \texttt{2.2.2} \cite{paszke2019pytorch}. A \texttt{NVIDIA GeForce RTX 3070} GPU is employed and the random seeds are set to 0 for all simulations, unless explicitly stated. The Adam optimizer \cite{kingma2014adam} is used for the training. Additionally, reference solutions are obtained using the finite element method (FEM), through an in-house implementation in \texttt{FreeFem++}  \cite{hecht2012new}.

\subsection{Laplace problem on L-shaped geometry}\label{sec:test1}
In the first test, we employ PIHKANs to solve the non-homogeneous Laplace problem in an L-shaped domain $\Omega \subset [-1,1]\times[-1,1]$. Specifically, we consider the equation $ -\nabla^2 u = 1 $ with homogeneous Dirichlet boundary conditions $u=0$. This problem serves as a benchmark, previously used in studies \cite{lu2021learning,mao2023physics,cho2024separable}, facilitating direct comparison with state-of-the-art methods. While generally straightforward, the problem is known to present challenges near the concave $3\pi/2$ angle.

The finite element method (FEM) is employed to compute the reference solution. A first-order discretization is used on a grid with average element size $h\approx 3\cdot 10^{-3}$. The PIHKAN network is composed by 5 inner layers, with 10 neurons each, and $P=5$. The initial learning rate is $5\cdot10^{-3}$. PIHNNs are also employed with the standard MLP architecture, where here we consider 5 inner layers with 30 neurons each. Training corresponds to 2000 epochs with 800 boundary training points for both. Being the problem not homogeneous, we leverage the method introduced in \Cref{eq:non-homo}, where the particular solution is given by $\mathcal{F}^{-1}[\bm{b}] = -(x^2+y^2)/4$.

The underlying problem is additionally implemented in the library DeepXDE \cite{Lu2021deep}, that we use to obtain the solution through classical PINNs. In this regard, we adopt the settings already provided by the library, i.e., 1200 interior training points, 120 boundary training points, 4 inner layers with 50 neurons each, $5\cdot 10^4$ epochs
and initial learning rate $10^{-3}$.

\begin{figure}[!ht]
        \centering
        \begin{subfigure}[t]{0.4\textwidth}
            \centering
            \includegraphics[width=\linewidth]{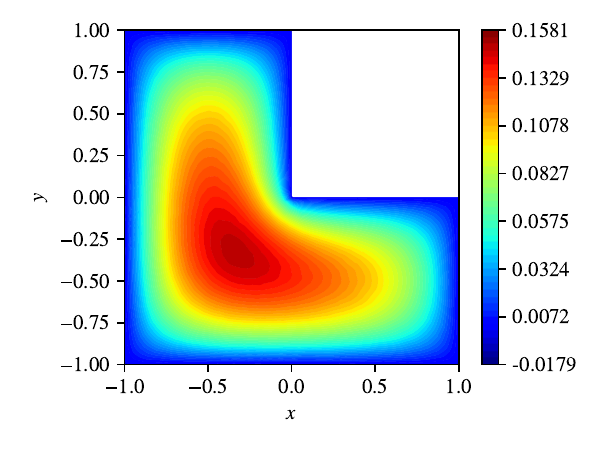}
            \caption{FEM.}
        \end{subfigure}
        \begin{subfigure}[t]{0.4\textwidth}
            \centering
            \includegraphics[width=\linewidth]{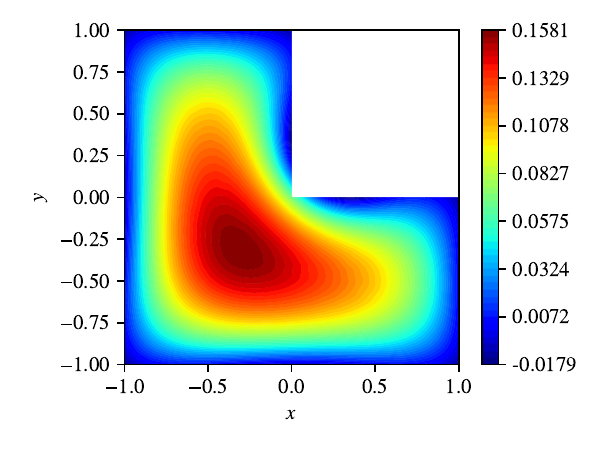}
            \caption{PINN.}
        \end{subfigure}
        \\
        \begin{subfigure}[t]{0.4\textwidth}
            \centering
            \includegraphics[width=\linewidth]{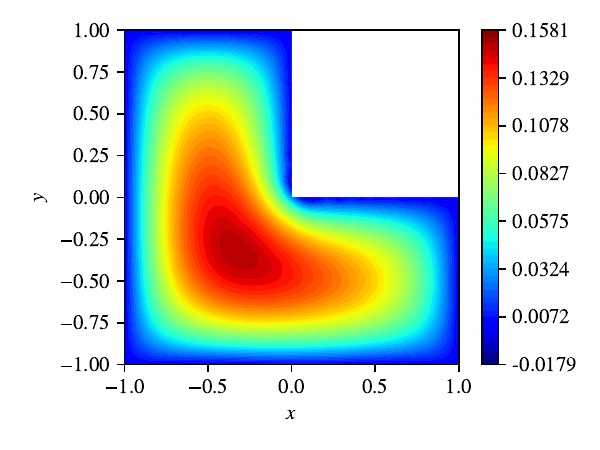}
            \caption{PIHNN.}
        \end{subfigure}
        \begin{subfigure}[t]{0.4\textwidth}
            \centering
            \includegraphics[width=\linewidth]{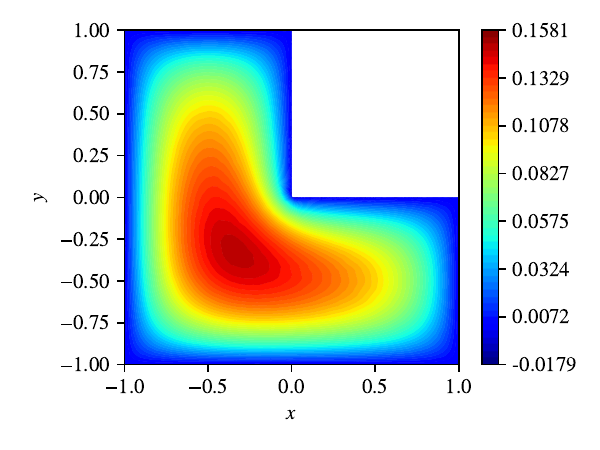}
            \caption{PIHKAN.}
        \end{subfigure}
        \caption{Contour plots of the various solutions from \Cref{sec:test1}. The color represents the value of $u$ and the same color scale is applied to all plots.}
        \label{fig:test1}
    \end{figure}

We show the contour plots of all solutions in \Cref{fig:test1}.

\begin{figure}[!ht]
        \centering
        \begin{subfigure}[t]{0.32\textwidth}
            \centering
            \includegraphics[width=\linewidth]{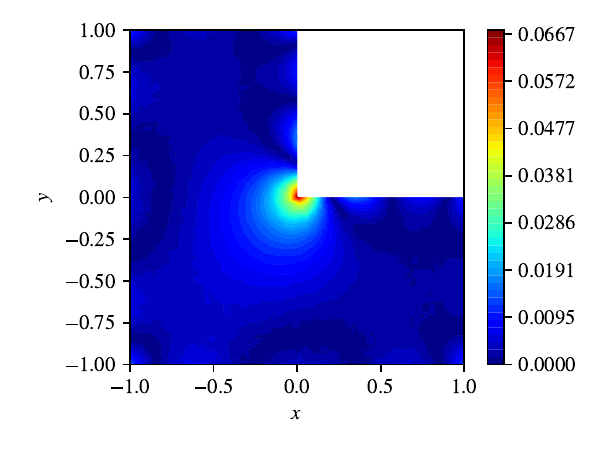}
            \caption{PINN.}
        \end{subfigure}
        \begin{subfigure}[t]{0.32\textwidth}
            \centering
            \includegraphics[width=\linewidth]{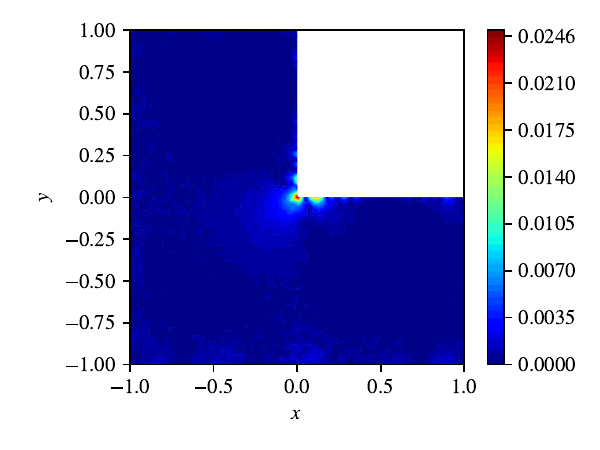}
            \caption{PIHNN.}
        \end{subfigure}
        \begin{subfigure}[t]{0.32\textwidth}
            \centering
            \includegraphics[width=\linewidth]{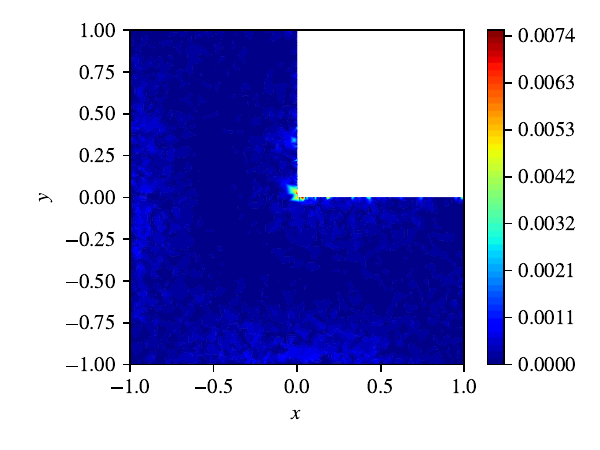}
            \caption{PIHKAN.}
        \end{subfigure}
        \caption{Errors $|u-\tilde{u}|$ of the various solutions from \Cref{sec:test1} with respect to the reference FEM solution. Different color ranges are adopted.}
        \label{fig:test1error}
\end{figure}

\begin{figure}[!ht]
        \centering
        \begin{subfigure}[t]{0.4\textwidth}
            \centering
            \includegraphics[width=\linewidth]{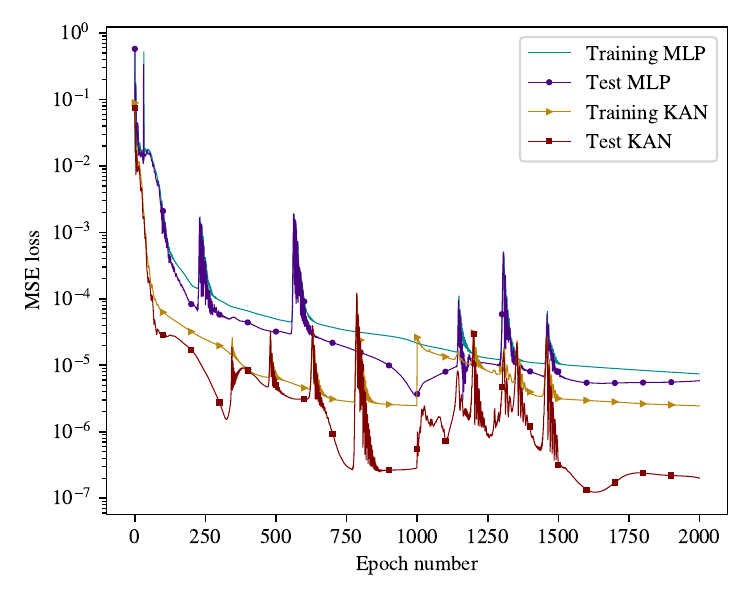}
            \caption{Laplace test.}
        \end{subfigure}
        \begin{subfigure}[t]{0.4\textwidth}
            \centering
            \includegraphics[width=\linewidth]{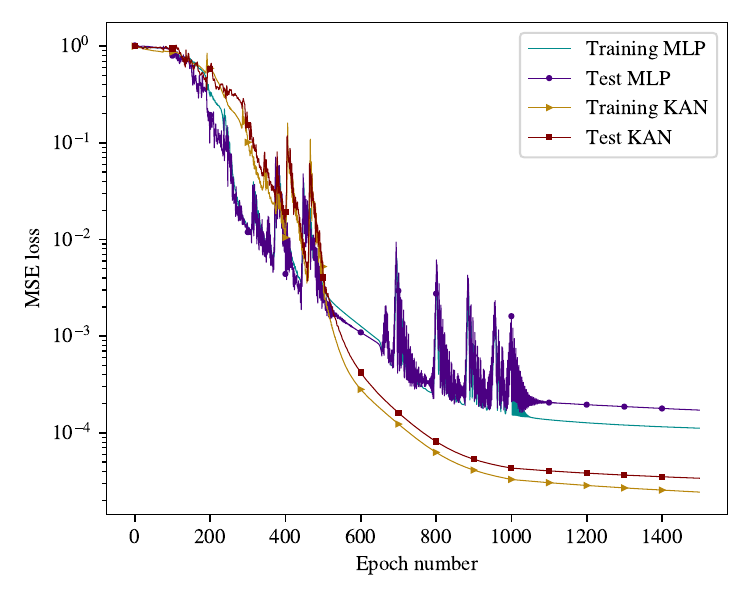}
            \caption{Helmholtz test.}
        \end{subfigure}
        \caption{Training and test loss curves for the experiments in \Cref{sec:test1} and \Cref{sec:test2}.}
        \label{fig:loss}
\end{figure}

A preliminary qualitative analysis suggests that classical PINNs exhibit lower accuracy compared to both holomorphic methods. This can be observed, for example, in the central region and near the concave angle. However, to provide a more objective comparison, we compute the errors relative to the FEM solution. Specifically, we define the error as the relative $L^2(\Omega)$ error:  

\begin{equation*}
    \text{Error}(\tilde{u},u) := \sqrt{ \frac{\int_\Omega (u-\tilde{u})^2} {\int_{\Omega} u^2 },}
\end{equation*}  

where the integrals are approximated using Monte Carlo sampling with $10^4$ uniformly distributed interior points. Errors relative to the FEM solution are presented in \Cref{tab:test1}, alongside the corresponding training times and the number of network parameters. Additionally, \Cref{fig:test1error} illustrates that the error is predominantly concentrated around the critical corner, while \Cref{fig:loss} demonstrates the improved convergence properties of PIHKANs.

\begin{table}[ht]
    \centering
    \bgroup
    \begin{tabular}{ c c c c }
        \toprule Method & Training time (s) & \#parameters & Relative $L^2(\Omega)$ error \\
        \midrule \midrule
        PINN & 526 & 7851 & $1.18\cdot 10^{-1}$  \\
        \midrule
        PIHNN & 13 & 7622 & $2.77\cdot 10^{-2}$  \\
        \midrule
        PIHKAN & 12 & 4302 & $1.76\cdot 10^{-2}$  \\
        \bottomrule
    \end{tabular}
    \egroup
    \caption{Accuracy and efficiency comparison between the different methods employed in \Cref{sec:test1}. The second column provides the number of trainable real-valued network parameters. Errors are computed with respect to the FEM solution.}
    \label{tab:test1}
\end{table}

It can be readily noticed that holomorphic methods outperform classical PINNs, both in terms of accuracy and efficiency, as already expected from \citet{calafa2024}. The comparison between PIHNN and PIHKAN, however, is more nuanced: while both achieve similar accuracy and performance, PIHNN employs more memory. This aligns with other studies that suggest KANs typically require smaller nets \cite{shukla2024,carneros2024comparison}. 

In \Cref{tab:ablation}, we present an ablation study in which we conduct multiple simulations by varying the number of layers, the number of neurons per layer, the polynomial degree $P$, and the random seed. The results demonstrate that the method remains stable under all variations. It is important to note that the training times are nearly identical, suggesting that the computational bottleneck is primarily due to the high number of epochs, rather than the network evaluation, due to the reduced number of training points and weights.

\begin{table}[ht]
    \centering
    \begin{tabular}{c c c c c c c }
        \toprule \#layers & \#neurons & $P$ & Seed & Training time (s) & \#parameters & Relative $L^2(\Omega)$ error \\
        \midrule \midrule
        5 & 10 & 5 & 0 & 12 & 4302 & $1.76 \cdot 10^{-2}$ \\
        \midrule
        6 & 10 & 5 & 0 & 13 & 5322 & $1.22 \cdot 10^{-2}$ \\
        \midrule
        5 & 12 & 5 & 0 & 12 & 6122 & $1.25 \cdot 10^{-2}$ \\
        \midrule
        5 & 10 & 6 & 0 & 12 & 5142 & $1.64 \cdot 10^{-2}$ \\
        \midrule
        5 & 10 & 5 & 1 & 12 & 4302 & $1.85 \cdot 10^{-2}$ \\
        \midrule
        5 & 10 & 5 & 42 & 12 & 4302 & $1.42 \cdot 10^{-2}$ \\
        \bottomrule
    \end{tabular}
    \caption{PIHKAN ablation study for the test in \Cref{sec:test1}.}
    \label{tab:ablation}
\end{table}

The plots in \Cref{fig:test1error} indicate that adaptive sampling could significantly reduce the solution error by concentrating training points around the critical corner. Wu et al.~\cite{wu2023comprehensive} proposed residual-based adaptive distribution (RAD) as an effective method for adaptive re-sampling, utilizing residual error information.
The process begins with training on an initial batch of points for a reduced number of epochs to obtain a temporary solution, denoted as $\tilde{\varphi}_1, \dots, \tilde{\varphi}_{N^*}$. The corresponding residuals are then computed as  

\begin{equation}\label{eq:epsilon}
\varepsilon := \mathcal{B}[\mathcal{F}^*[\tilde{\varphi}_1, \dots, \tilde{\varphi}_{N^*}]].
\end{equation}

Evaluating $\varepsilon$ on a refined set of boundary points helps identify sensitive areas where the network exhibits lower accuracy. This information guides the re-sampling of training points according to the probability distribution  

\begin{equation*}
\mathcal{P} \propto \frac{\varepsilon^{k_{rad}}}{\mathbb{E}[\varepsilon^{k_{rad}}]} + c_{rad},
\end{equation*}

where $k_{rad}$ and $c_{rad}$ are constants, set to $k_{rad} = c_{rad} = 1$ as recommended in the original work.

It should be noted that \Cref{eq:epsilon} yields in general a vectorial residual as boundary conditions could be hybrid and therefore $\mathcal{B}$ multi-dimensional. In order to take into account the heterogeneity of boundary conditions and hyperparameters $\bm{\lambda}_\mathcal{B}$, the method is applied separately for each dimension of $\mathcal{B}$.

The novelty of this study lies in applying RAD to boundary conditions, rather than to PDE residuals. Our objective is to demonstrate the continued effectiveness within this new context.

To evaluate boundary-based RAD, we initially sample 800 training points uniformly from $\partial\Omega$. A PIHKAN architecture from the experiment above, comprising $P=5$, 5 inner layers, and 10 neurons per layer, undergoes 1000 epochs of training with an initial learning rate of $10^{-2}$. Subsequently, RAD is employed with a fine grid of $10^4$ points to sample 800 new training points. Network training then continues on this new batch for an additional 1000 epochs. These settings largely mirror those in Table \ref{tab:test1}. Test results are presented in Table \ref{tab:RAD}, comparing RAD sampling against a standard training process, both initialized with random seed 0.

\begin{table}[ht]
    \centering
    \bgroup
    \begin{tabular}{ c c c c c }
        \toprule Method & Adaptive sampling & Training time (s) & \#parameters & Relative $L^2(\Omega)$ error \\
        \midrule \midrule
        PIHKAN & No & 12 & 4302 & $1.76\cdot 10^{-2}$  \\
        \midrule
        PIHKAN & Yes & 13 & 4302 & $7.13\cdot 10^{-3}$  \\
        \bottomrule
    \end{tabular}
    \egroup
    \caption{Evaluation of the benefits from the RAD adaptive sampling \cite{wu2023comprehensive}. RAD leads to negligible additional costs and effectively improves the accuracy of the learned solution.}
    \label{tab:RAD}
\end{table}

\begin{figure}[!ht]
        \centering
        \begin{subfigure}[t]{0.4\textwidth}
            \centering
            \includegraphics[width=\linewidth]{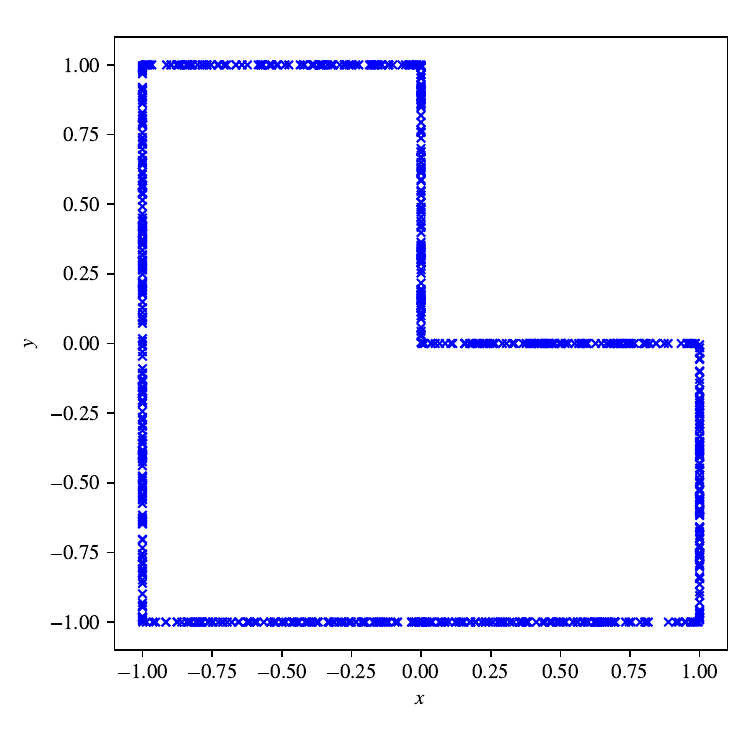}
            \caption{Uniform sampling.}
        \end{subfigure}
        \begin{subfigure}[t]{0.4\textwidth}
            \centering
            \includegraphics[width=\linewidth]{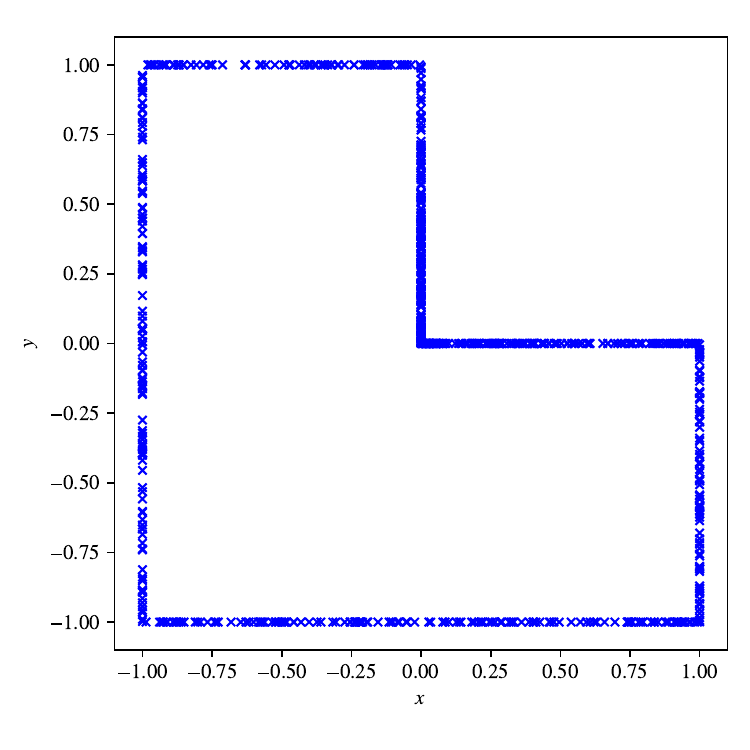}
            \caption{RAD sampling.}
        \end{subfigure}
        \caption{Comparison of the training points before (left) and after (right) the application of the boundary-based RAD in the test from \Cref{tab:RAD}.}
        \label{fig:bc_points}
    \end{figure}

The results demonstrate effective improvement without incurring significant additional computational costs. To elucidate the underlying reason for this enhancement, we visualize the training points in \Cref{fig:bc_points} both before (epoch 0) and after (epoch 1000) the application of RAD. As shown, the training points exhibit increased concentration around the critical corner, thereby compelling the network to more effectively learn the solution within this region.

\subsection{Helmholtz problem on a square}\label{sec:test2}
In this section, PIHNNs and PIHKANs are employed to solve the 2D Helmholtz equation. The method employs the Vekua operator, as defined in \Cref{eq:vekua}, incorporating a change of variable  $s=\sqrt{1-t}$. The resulting integral operator is

\begin{equation*}
    u(\mathbf{x}) = \text{Re}(\varphi(z)) - \int_0^1 \beta |z| J_1(\beta |z| s) \, \text{Re}(\varphi(z(1-s^2))) \, ds,
\end{equation*}
where $u$ is by construction the solution to the Helmholtz equation with wave number $\beta>0$, $\varphi$ is the output of a holomorphic network and we adopt the usual correspondence $\mathbf{x}=(\text{Re}(z),\text{Im}(z))$. The specific change of variable is motivated by its ability to enhance integration accuracy by eliminating the denominator term.

\begin{figure}[!ht]
        \centering
        \begin{subfigure}[t]{0.4\textwidth}
            \centering
            \includegraphics[width=\linewidth]{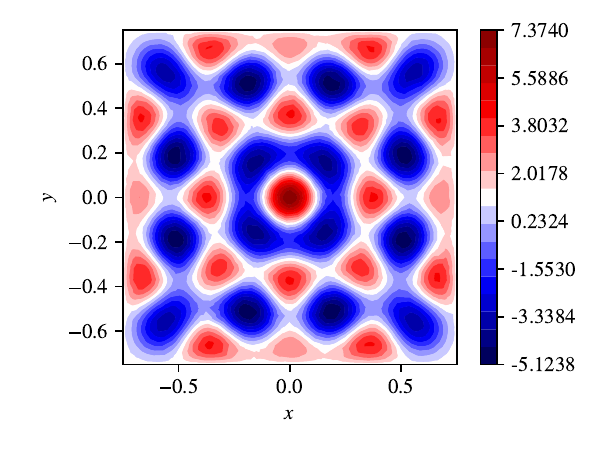}
            \caption{FEM.}
        \end{subfigure}
        \begin{subfigure}[t]{0.4\textwidth}
            \centering
            \includegraphics[width=\linewidth]{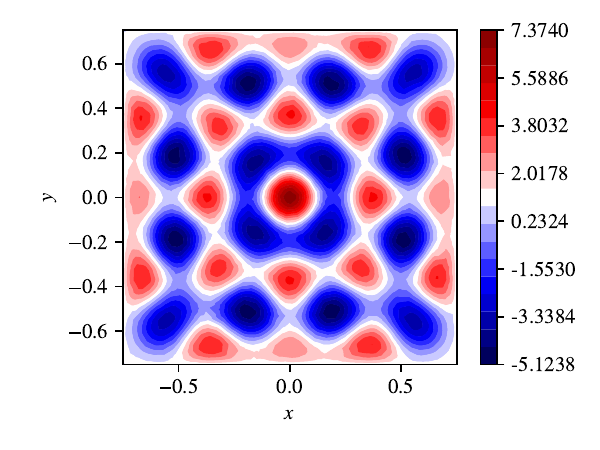}
            \caption{PINN.}
        \end{subfigure}
        \\
        \begin{subfigure}[t]{0.4\textwidth}
            \centering
            \includegraphics[width=\linewidth]{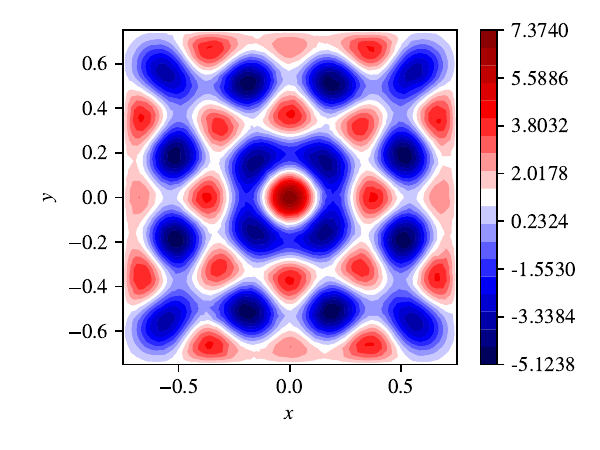}
            \caption{PIHNN.}
        \end{subfigure}
        \begin{subfigure}[t]{0.4\textwidth}
            \centering
            \includegraphics[width=\linewidth]{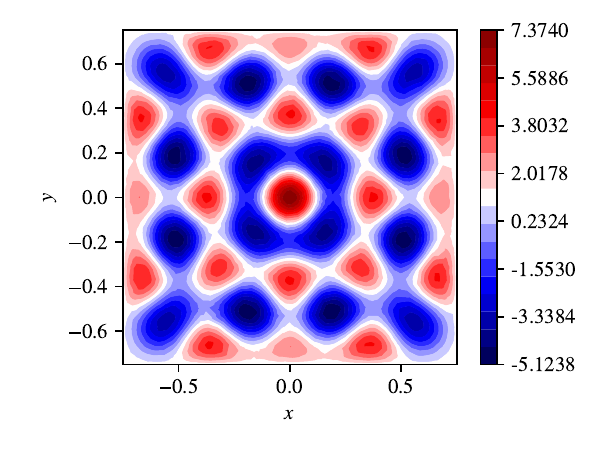}
            \caption{PIHKAN.}
        \end{subfigure}
        \caption{Contour plots of the various solutions from \Cref{sec:test2}. The color represents the value of $u$ and the same color scale is applied to all plots.}
        \label{fig:test2}
\end{figure}

\begin{figure}[!ht]
        \centering
        \begin{subfigure}[t]{0.32\textwidth}
            \centering
            \includegraphics[width=\linewidth]{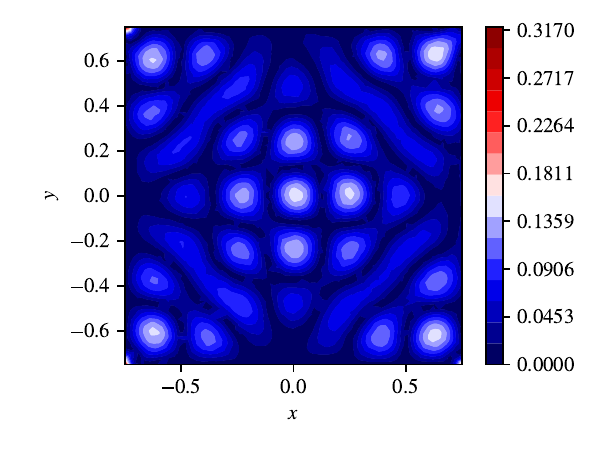}
            \caption{PINN.}
        \end{subfigure}
        \begin{subfigure}[t]{0.32\textwidth}
            \centering
            \includegraphics[width=\linewidth]{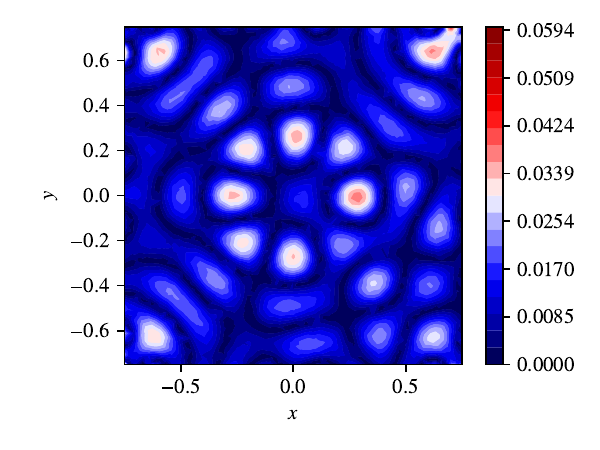}
            \caption{PIHNN.}
        \end{subfigure}
        \begin{subfigure}[t]{0.32\textwidth}
            \centering
            \includegraphics[width=\linewidth]{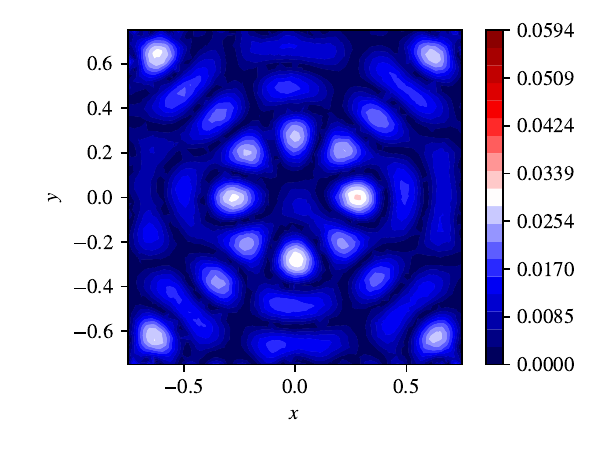}
            \caption{PIHKAN.}
        \end{subfigure}
        \caption{Errors $|u-\tilde{u}|$ of the various solutions from \Cref{sec:test2} with respect to the reference FEM solution. Different color ranges are adopted.}
        \label{fig:test2error}
\end{figure}

In our experiment, we employ 20 Gauss-Legendre quadrature points on the interval $[0,1]$ to calculate the Vekua integral. Furthermore, we consider the speed of sound in air $c=343$ m/s and a sound frequency of $f=1000$ Hz, which correspond to a wave number $\beta \approx 18$~m$^{-1}$. The Helmholtz problem is solved on a square of length 1.5 m with Dirichlet boundary conditions $u=1$ applied uniformly across the boundary. The adopted PIHKAN consists of 2 hidden layers, each with 10 neurons, $P=4$, 1500 epochs, 600 boundary training points and learning rate $10^{-2}$. We compare the KAN architecture against a PIHNN where the number of neurons per layer is increased to 40. In addition, we include the results from a classical PINN implemented in the library DeepXDE. In particular, we adopt some default settings: 4 hidden layers with 50 neurons each, $\sin(x)$ activation function, learning rate $10^{-3}$, $5\cdot 10^4$ epochs, 841 internal training points and 116 boundary training points. The contour plots can be seen in \Cref{fig:test2} and \Cref{fig:test2error}, where the reference FEM solution is obtained through a first-order discretization on a grid with average element size $h\approx 3 \cdot 10^{-3}$ m, i.e., far below the wavelength $\approx 0.34$ m.

Detailed information are given in \Cref{tab:test2}. In particular, the holomorphic approach is as expected superior to the classical physics-informed method. Furthermore, PIHKANs can achieve greater accuracy than PIHNN with a smaller network. 

On the other hand, the observed increase in training time signals a reduced efficiency in the evaluation of KAN activations. This phenomenon was not observed in \Cref{sec:test1}, due to the limited input and network size. In the current experiment, however, the input size is substantially larger, as each batch input is associated with 20 quadrature points. Consequently, the primary computational bottleneck in PIHKAN training lies in the slow evaluation of the trainable polynomials, specifically via the \texttt{torch.linalg.vander} function, which currently lacks CUDA optimization.

\begin{table}[ht]
    \centering
    \bgroup
    \begin{tabular}{c c c c }
        \toprule Method & Training time (s) & \#parameters & Relative $L^2(\Omega)$ error \\
        \midrule \midrule
        PINN & 534 & 7851 & $2.66\cdot 10^{-2}$  \\
        \midrule
        PIHNN & 13 & 3522 & $6.31\cdot 10^{-3}$  \\
        \midrule
        PIHKAN & 26 & 1002 & $4.89\cdot 10^{-3}$  \\
        \bottomrule
    \end{tabular}
    \egroup
    \caption{Accuracy and efficiency comparison between the different methods employed in \Cref{sec:test1}. The second column provides the number of trainable real-valued network parameters. Errors are computed with respect to the FEM solution.}
    \label{tab:test2}
\end{table}

\subsection{Linear elasticity on multiply-connected domain}\label{sec:test3}

\begin{figure}[!ht]
    \centering
    \begin{tikzpicture}[scale=1.01]
    \fill [blue!50!cyan] (0,0) rectangle (60pt,60pt);
    \fill [green!40!orange] (0,0) rectangle (-60pt,60pt);
    \fill [red!40!orange] (0,0) rectangle (-60pt,-60pt);
    \fill [yellow!40!orange] (0,0) rectangle (60pt,-60pt);
    \draw [] (-60pt,-60pt) rectangle (60pt,60pt);
    \draw [dotted] (-60pt,0) -- (-24pt,0);
    \draw [dotted] (60pt,0) -- (24pt,0);
    \draw [dotted] (0,-60pt) -- (0,-24pt);
    \draw [dotted] (0,60pt) -- (0,24pt);
    \draw [fill=white] (0,0) circle (24pt);
    \draw [fill=black] (0,0) circle (1pt);
    \node[] at (-4pt,-5pt) {$z_1$};
    \draw [<->] (2pt,2pt) -- (16pt,16pt);
    \draw [<->] (-59pt,-65pt) -- (59pt,-65pt);
    \node[] at (5pt,12pt) {$r$};
    \node[] at (0pt,-71pt) {$L$};
    \draw [<-] (-80pt,0) -- (-62pt,0);
    \draw [<-] (-80pt,20pt) -- (-62pt,20pt);
    \draw [<-] (-80pt,40pt) -- (-62pt,40pt);
    \draw [<-] (-80pt,-20pt) -- (-62pt,-20pt);
    \draw [<-] (-80pt,-40pt) -- (-62pt,-40pt);
    \phantom{\draw [->] (0,-80pt) -- (0,-62pt);}
    \draw [<-] (80pt,0) -- (62pt,0);
    \draw [<-] (80pt,20pt) -- (62pt,20pt);
    \draw [<-] (80pt,40pt) -- (62pt,40pt);
    \draw [<-] (80pt,-20pt) -- (62pt,-20pt);
    \draw [<-] (80pt,-40pt) -- (62pt,-40pt);
    \phantom{\draw [->] (0,80pt) -- (0,62pt);}
    \node[] at (-74pt,6pt) {$\bm{t}_0$};
    \node[] at (74pt,6pt) {$\bm{t}_0$};
    \phantom{\node[] at (0pt,-83pt) {$\bm{t}_0$};}
    \end{tikzpicture}
    \caption{Geometry for the test in \Cref{sec:test3}. A uniaxial tension with unitary intensity $|\bm{t}_0|=1$ is applied on a square plate of length $L=2.5$ containing an inner circle of radius $r=1$. The four domains employed in the domain decomposition are displayed with different colors. The Laurent method employs only one singularity net, represented by the point $z_1$ at the center of the plate.}
    \label{fig:test3geometry}
\end{figure}
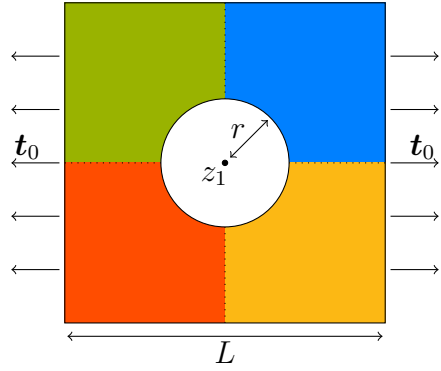

We test a multiply-connected domain to assess the effectiveness of the method proposed in \Cref{sec:laurent}. Specifically, we replicate the benchmark described by \citet[Section 4.1.2]{calafa2024}, which solves the linear elasticity problem on a square plate of length $L=2.5$ containing an inner circular hole of radius $r=1$, subjected to unit uniaxial tension (\Cref{fig:test3geometry}). A domain decomposition strategy was therein applied to PIHNNs as a mean to extend the original applicability to multiply-connected domains. In this work, we apply the Laurent method to holomorphic networks, using both MLP and KAN variants, and compare it with the original domain decomposition strategy. A direct comparison with standard PINNs is instead omitted, as this was already performed in \citet{calafa2024}, where holomorphic networks achieved higher accuracy over PINNs and approximately a 60-times reduction in training time.

The domain decomposition is applied to the four quadrants, as displayed in \Cref{fig:test3geometry}. The MLP network is composed by 2 hidden layers with 15 neurons each, whereas the KAN network has 5 neurons per layer. The Laurent method is also employed on single-domain holomorphic MLPs and KANs, where the former has now 20 neurons and the latter has 7 neurons each for both the analytic ($\varphi_{n,0}$) and singular ($\varphi_{n,1}$) sub-networks. The formula in \Cref{eq:laurent} is employed, by choosing $z_s$ for $s=1$ at the origin of the coordinate system. An initial learning rate of $10^{-2}$ is used, with 600 training points uniformly distributed on the boundary and $4\cdot 10^3$ epochs. Finally, we select $P=4$ for the KAN networks.

\begin{figure}[!ht]
        \centering
        \begin{subfigure}[t]{0.29\textwidth}
            \centering
            \includegraphics[width=\linewidth]{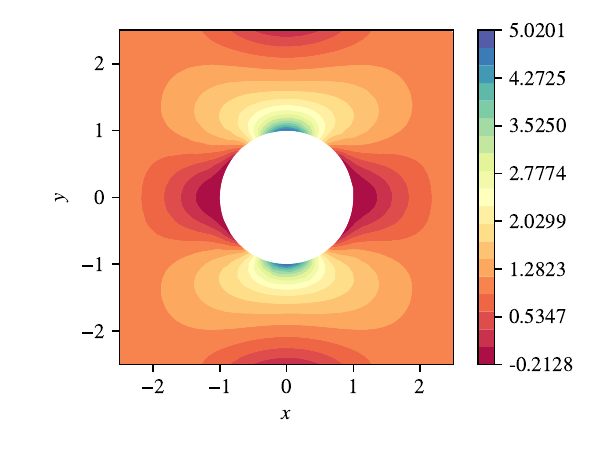}
            \caption{FEM ($\sigma_{xx}$).}
        \end{subfigure}
        \begin{subfigure}[t]{0.29\textwidth}
            \centering
            \includegraphics[width=\linewidth]{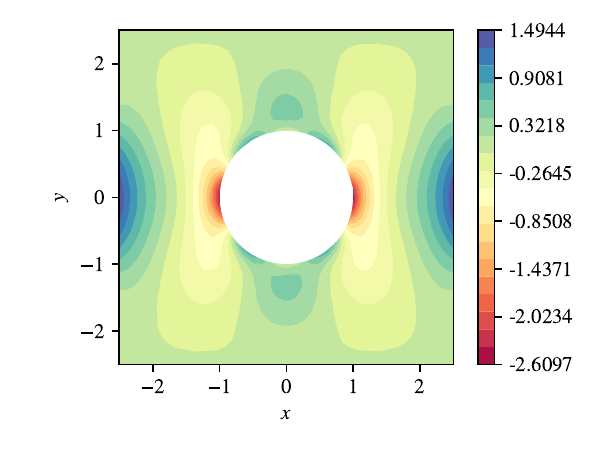}
            \caption{FEM ($\sigma_{yy}$).}
        \end{subfigure}
        \begin{subfigure}[t]{0.29\textwidth}
            \centering
            \includegraphics[width=\linewidth]{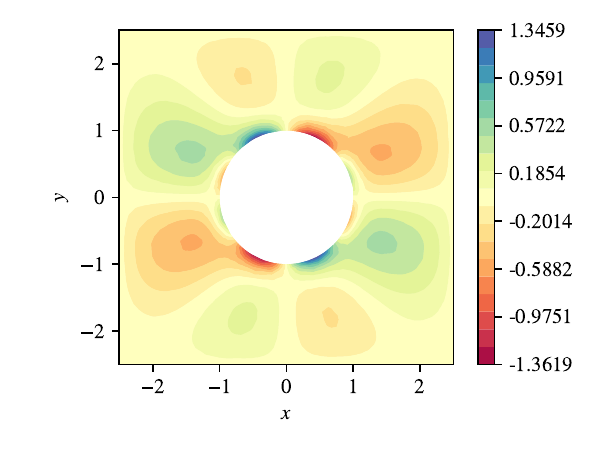}
            \caption{FEM ($\sigma_{xy})$.}
        \end{subfigure}
        \\
        \centering
        \begin{subfigure}[t]{0.29\textwidth}
            \centering
            \includegraphics[width=\linewidth]{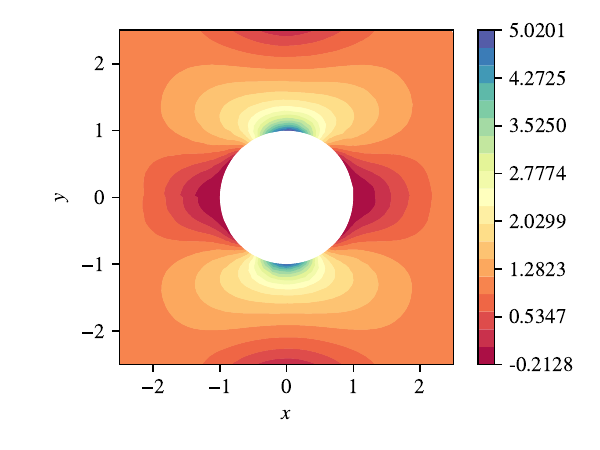}
            \caption{DD-PIHNN ($\sigma_{xx}$).}
        \end{subfigure}
        \begin{subfigure}[t]{0.29\textwidth}
            \centering
            \includegraphics[width=\linewidth]{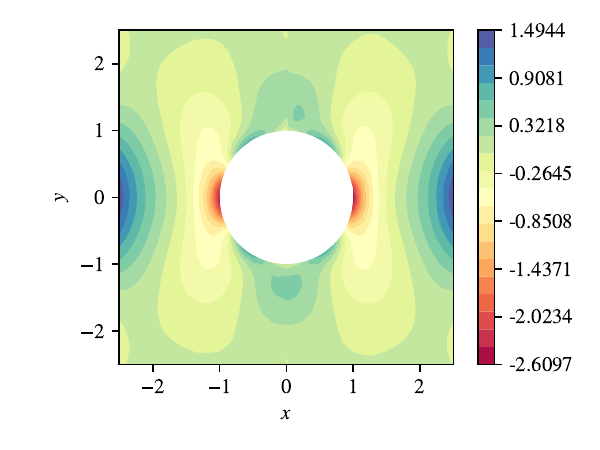}
            \caption{DD-PIHNN ($\sigma_{yy}$).}
        \end{subfigure}
        \begin{subfigure}[t]{0.29\textwidth}
            \centering
            \includegraphics[width=\linewidth]{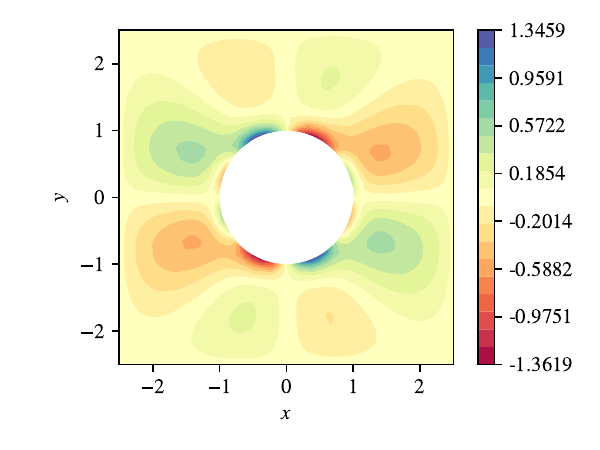}
            \caption{DD-PIHNN ($\sigma_{xy}$).}
        \end{subfigure}
        \\
        \centering
        \begin{subfigure}[t]{0.29\textwidth}
            \centering
            \includegraphics[width=\linewidth]{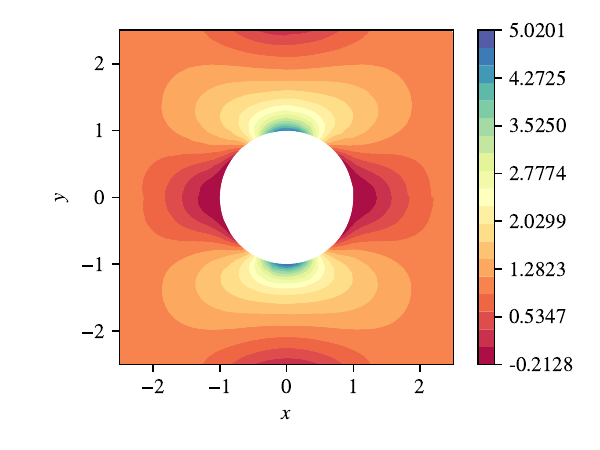}
            \caption{DD-PIHKAN ($\sigma_{xx}$).}
        \end{subfigure}
        \begin{subfigure}[t]{0.29\textwidth}
            \centering
            \includegraphics[width=\linewidth]{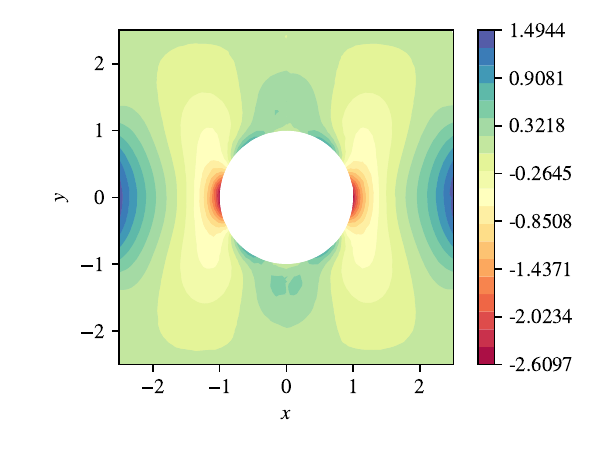}
            \caption{DD-PIHKAN ($\sigma_{yy}$).}
        \end{subfigure}
        \begin{subfigure}[t]{0.29\textwidth}
            \centering
            \includegraphics[width=\linewidth]{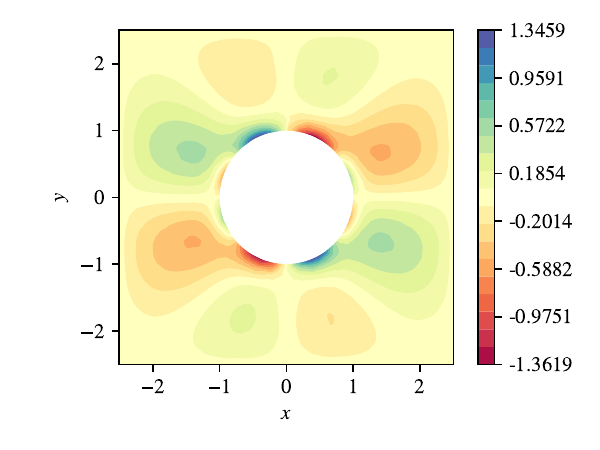}
            \caption{DD-PIHKAN ($\sigma_{xy})$.}
        \end{subfigure}
        \\
        \centering
        \begin{subfigure}[t]{0.29\textwidth}
            \centering
            \includegraphics[width=\linewidth]{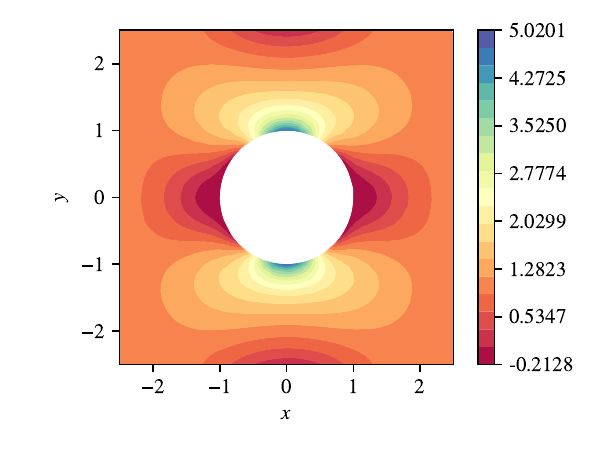}
            \caption{L-PIHNN ($\sigma_{xx}$).}
        \end{subfigure}
        \begin{subfigure}[t]{0.29\textwidth}
            \centering
            \includegraphics[width=\linewidth]{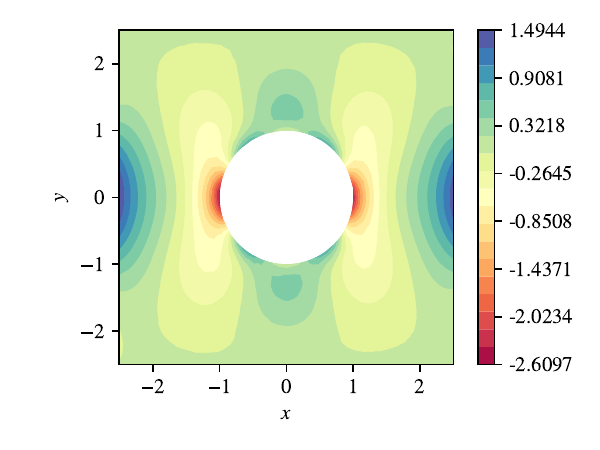}
            \caption{L-PIHNN ($\sigma_{yy}$).}
        \end{subfigure}
        \begin{subfigure}[t]{0.29\textwidth}
            \centering
            \includegraphics[width=\linewidth]{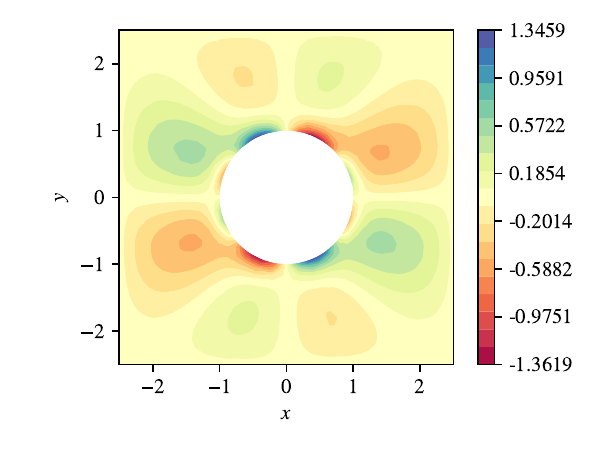}
            \caption{L-PIHNN ($\sigma_{xy}$).}
        \end{subfigure}
        \\
        \centering
        \begin{subfigure}[t]{0.29\textwidth}
            \centering
            \includegraphics[width=\linewidth]{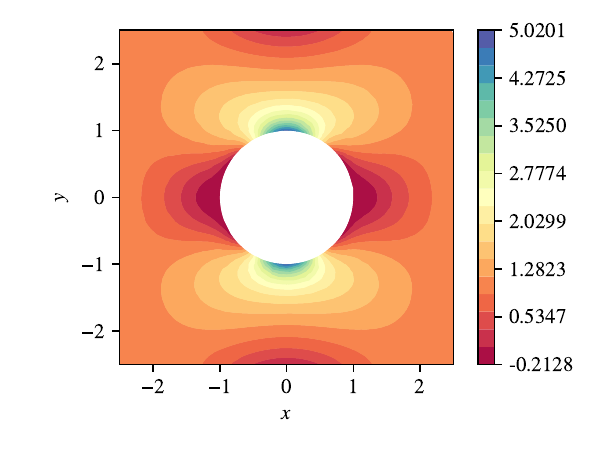}
            \caption{L-PIHKAN ($\sigma_{xx}$).}
        \end{subfigure}
        \begin{subfigure}[t]{0.29\textwidth}
            \centering
            \includegraphics[width=\linewidth]{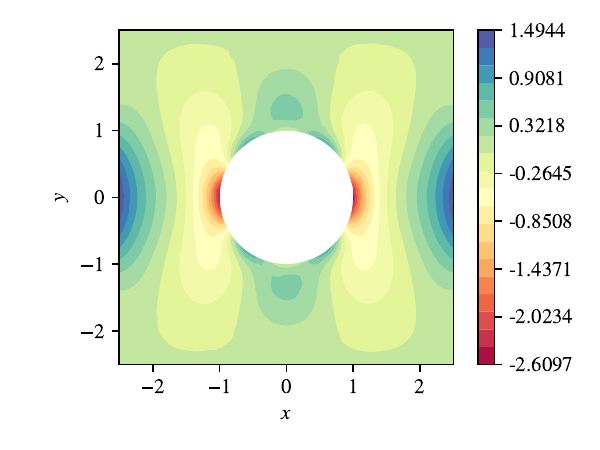}
            \caption{L-PIHKAN ($\sigma_{yy}$).}
        \end{subfigure}
        \begin{subfigure}[t]{0.29\textwidth}
            \centering
            \includegraphics[width=\linewidth]{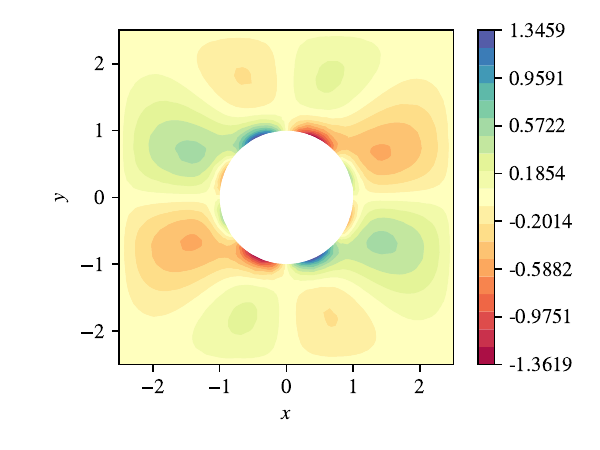}
            \caption{L-PIHKAN ($\sigma_{xy}$).}
        \end{subfigure}
        \caption{Contour plots of the various solutions from \Cref{sec:test3}. The color represents the value of the stress component and the same color scale is applied to all methods.}
        \label{fig:test3}
\end{figure}

\begin{figure}[!ht]
        \centering
        \begin{subfigure}[t]{0.29\textwidth}
            \centering
            \includegraphics[width=\linewidth]{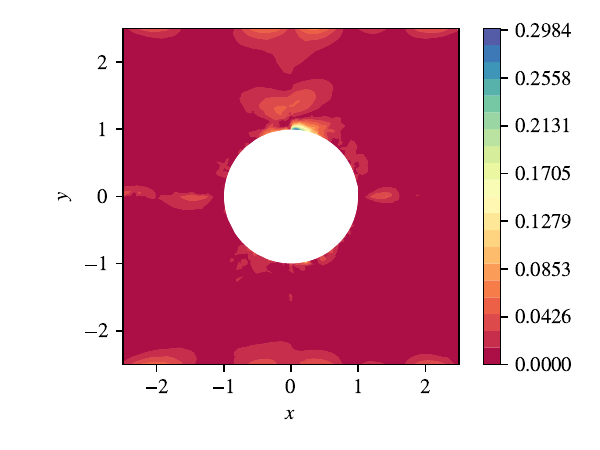}
            \caption{DD-PIHNN ($\sigma_{xx}$).}
        \end{subfigure}
        \begin{subfigure}[t]{0.29\textwidth}
            \centering
            \includegraphics[width=\linewidth]{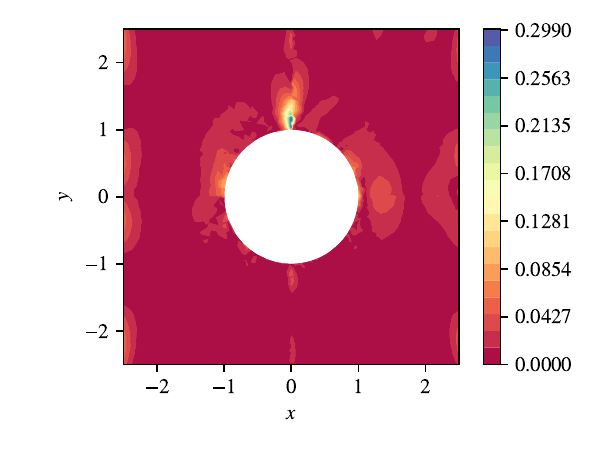}
            \caption{DD-PIHNN ($\sigma_{yy}$).}
        \end{subfigure}
        \begin{subfigure}[t]{0.29\textwidth}
            \centering
            \includegraphics[width=\linewidth]{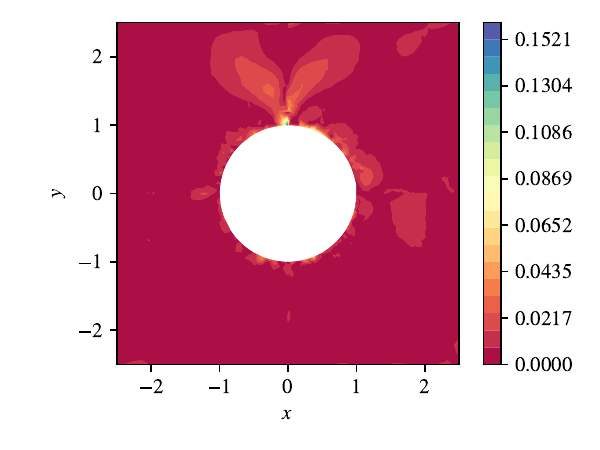}
            \caption{DD-PIHNN ($\sigma_{xy}$).}
        \end{subfigure}
        \\
        \centering
        \begin{subfigure}[t]{0.29\textwidth}
            \centering
            \includegraphics[width=\linewidth]{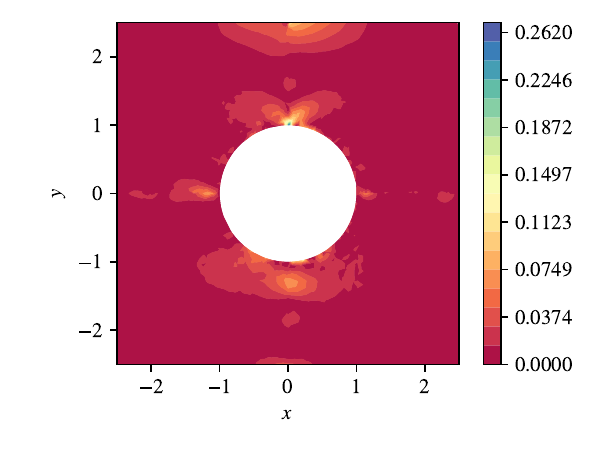}
            \caption{DD-PIHKAN ($\sigma_{xx}$).}
        \end{subfigure}
        \begin{subfigure}[t]{0.29\textwidth}
            \centering
            \includegraphics[width=\linewidth]{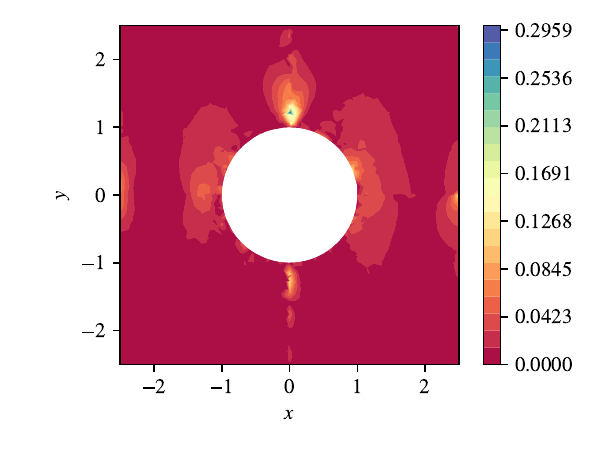}
            \caption{DD-PIHKAN ($\sigma_{yy}$).}
        \end{subfigure}
        \begin{subfigure}[t]{0.29\textwidth}
            \centering
            \includegraphics[width=\linewidth]{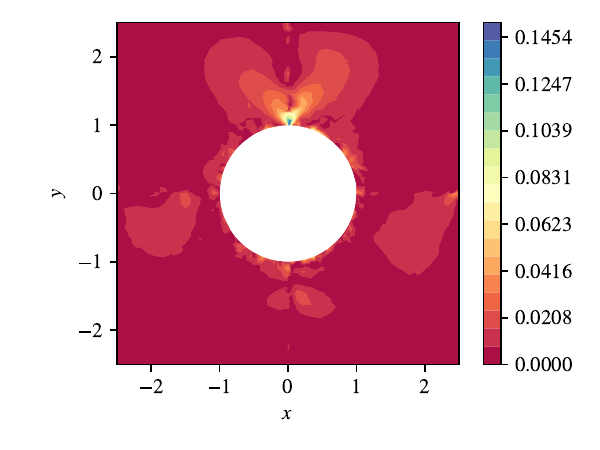}
            \caption{DD-PIHKAN ($\sigma_{xy})$.}
        \end{subfigure}
        \\
        \centering
        \begin{subfigure}[t]{0.29\textwidth}
            \centering
            \includegraphics[width=\linewidth]{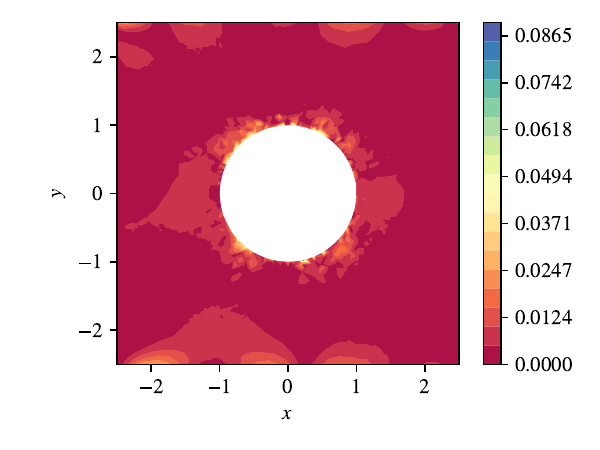}
            \caption{L-PIHNN ($\sigma_{xx}$).}
        \end{subfigure}
        \begin{subfigure}[t]{0.29\textwidth}
            \centering
            \includegraphics[width=\linewidth]{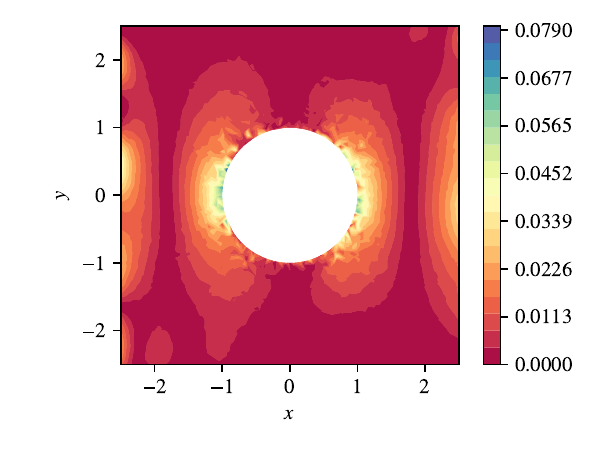}
            \caption{L-PIHNN ($\sigma_{yy}$).}
        \end{subfigure}
        \begin{subfigure}[t]{0.29\textwidth}
            \centering
            \includegraphics[width=\linewidth]{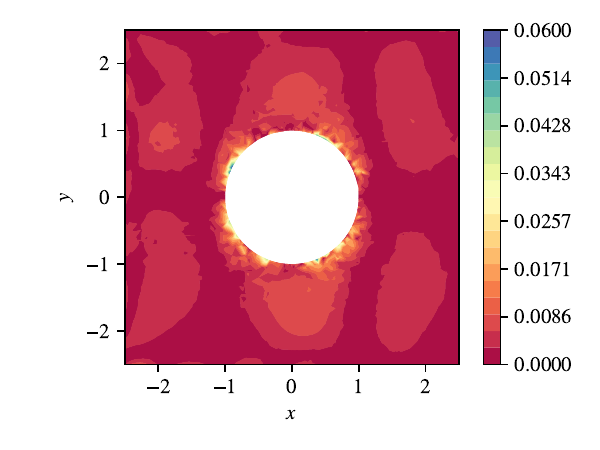}
            \caption{L-PIHNN ($\sigma_{xy}$).}
        \end{subfigure}
        \\
        \centering
        \begin{subfigure}[t]{0.29\textwidth}
            \centering
            \includegraphics[width=\linewidth]{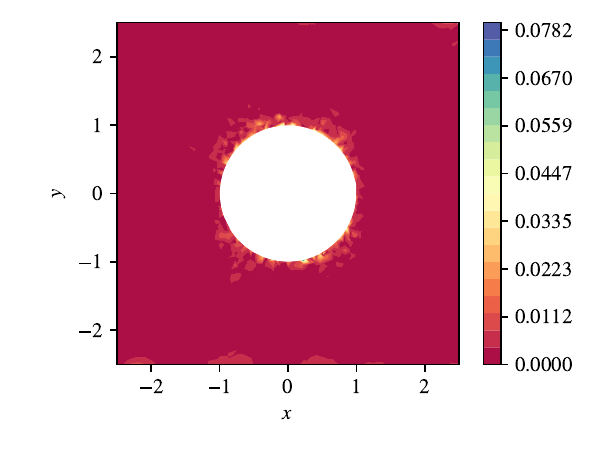}
            \caption{L-PIHKAN ($\sigma_{xx}$).}
        \end{subfigure}
        \begin{subfigure}[t]{0.29\textwidth}
            \centering
            \includegraphics[width=\linewidth]{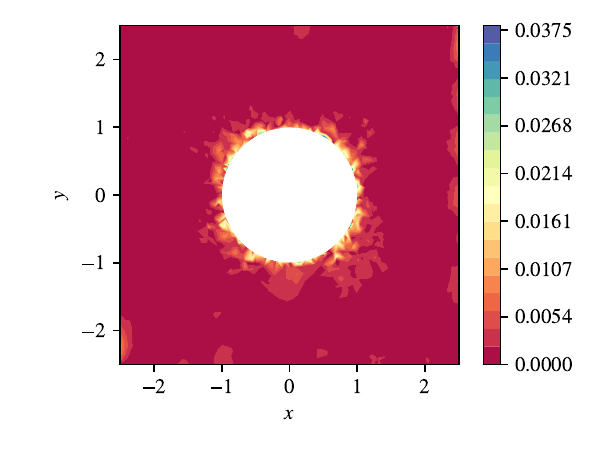}
            \caption{L-PIHKAN ($\sigma_{yy}$).}
        \end{subfigure}
        \begin{subfigure}[t]{0.29\textwidth}
            \centering
            \includegraphics[width=\linewidth]{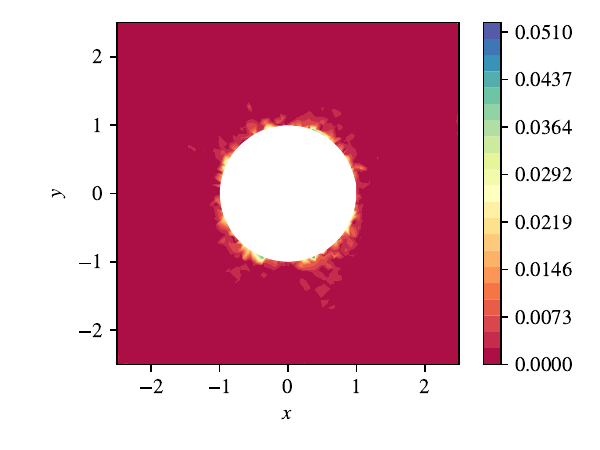}
            \caption{L-PIHKAN ($\sigma_{xy}$).}
        \end{subfigure}
        \caption{Components of the error $|\bm{\upsigma}-\tilde{\bm{\upsigma}}|$ from \Cref{sec:test3} with respect to the reference FEM solution.}
        \label{fig:test3error}
\end{figure}

A finite element method with second order elements of length approximately $0.04$ is employed to evaluate the accuracy of the learned solutions. As anticipated in \Cref{sec:methods}, linear elasticity is solved through the Kolosov-Muskhelishvili representation

\begin{equation*}
\begin{cases}
u_x &= \frac{1}{2\mu} \text{Re}\left(\kappa\varphi_1 - z \overline{\varphi_1'} - \overline{\varphi_2}\right), \\
u_y &= \frac{1}{2\mu} \text{Im}\left(\kappa\varphi_1 - z \overline{\varphi_1'} - \overline{\varphi_2}\right), \\
\sigma_{xx} &= \text{Re}\left(2 \varphi'_1 - \overline{z} \varphi_1'' -\varphi'_2\right), \\
\sigma_{yy} &= \text{Re}\left(2 \varphi'_1 + \overline{z} \varphi_1'' +\varphi'_2\right), \\
\sigma_{xy} &= \text{Im}\left(\overline{z} \varphi_1'' +\varphi'_2\right),
\end{cases}
\end{equation*}
where $\varphi_1,\varphi_2$ are the outputs of two holomorphic neural networks. Furthermore, the continuity condition at the domain interface is assigned as
\begin{equation*}
    \left[\bm{u}\right] = \left[\bm{\upsigma} \right] \cdot \bm{n} = \bm{0},
\end{equation*}
where $\left[\cdot\right]$ denotes the discontinuous jump between the two sides of the interface, $\bm{n}$ is the normal unit vector to the interface, $\bm{u}=[u_x,u_y]$ is the displacement vector, $\upsigma$ is the stress tensor with normal components $\sigma_{xx},\sigma_{yy}$ and shear component $\sigma_{xy}$. We address the reader to previous works \cite{calafa2024,calafa2025solving} for further details on linear elasticity solvers through holomorphic neural networks.

Learned solutions can be visually compared in \Cref{fig:test3} and \Cref{fig:test3error}, where the Laurent-based methods demonstrate to reach better convergence, especially in correspondence of the four axes.

Furthermore, \Cref{tab:test3} includes more details on the relative $L^2$ errors with respect to the three components of the stress tensor and the training times.

\begin{table}[ht]
    \centering
    \bgroup
    \begin{tabular}{c c c c c c }
        \toprule \multirow{2}{*}{Method} & \multirow{2}{*}{Training time (s)}  & \multirow{2}{*}{\#parameters}& \multicolumn{3}{c}{Relative $L^2(\Omega)$ error}\\
         & & & $\sigma_{xx}$ & $\sigma_{yy}$ & $\sigma_{xy}$  \\ 
        \midrule \midrule
        DD-PIHNN & 89 & 4576 & $1.41\cdot 10^{-2}$ & $4.70\cdot 10^{-2}$ & $3.66\cdot 10^{-2}$ \\
        \midrule
        DD-PIHKAN & 114 & 2416 & $1.32\cdot 10^{-2}$& $4.35\cdot 10^{-2}$& $4.00\cdot 10^{-2}$\\
        \midrule
        L-PIHNN & 117 & 3850 & $5.61\cdot 10^{-3}$ & $2.92\cdot 10^{-2}$ &$2.43\cdot 10^{-2}$\\
        \midrule
        L-PIHKAN & 195 & 2138 & $3.85\cdot 10^{-3}$ & $8.43\cdot 10^{-3}$ &$1.44\cdot 10^{-2}$\\
        \bottomrule
    \end{tabular}
    \egroup
    \caption{Accuracy and efficiency comparison between the different methods employed in \Cref{sec:test3}. The second column provides the number of trainable real-valued network parameters. Errors are computed with respect to the FEM solution.}
    \label{tab:test3}
\end{table}

The use of the Laurent method demonstrates higher accuracy compared to domain decomposition. Specifically, the Laurent-based PIHKAN achieves the lowest errors with the smallest neural network size, but it comes at the cost of longer computation times. This is due to two main reasons: first, the test in \Cref{sec:test2} has already shown that KAN activations require more time to evaluate, partly due to the suboptimal CUDA implementation in \texttt{PyTorch}. Second, the code implementation of the domain decomposition approach is fully parallelized, with four domains handled by stacked neural networks, whereas the singular networks $\{\varphi_{n,s}\}_{s=1}^S$ in the Laurent method are evaluated sequentially, which slows down training but also suggests potential for optimization. 

It was also noted that managing four domains makes the training process in domain decomposition more stochastic and prone to instabilities, as well as more sensitive to hyperparameter tuning, compared to the Laurent-based method.

\begin{figure}[!ht]
        \centering
        \begin{subfigure}[t]{0.4\textwidth}
            \centering
            \includegraphics[width=\linewidth]{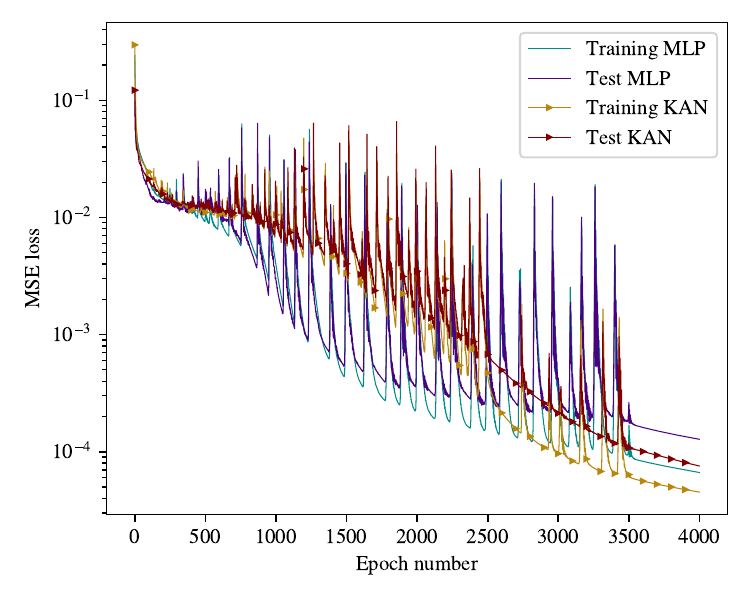}
            \caption{Domain decomposition.}
        \end{subfigure}
        \begin{subfigure}[t]{0.4\textwidth}
            \centering
            \includegraphics[width=\linewidth]{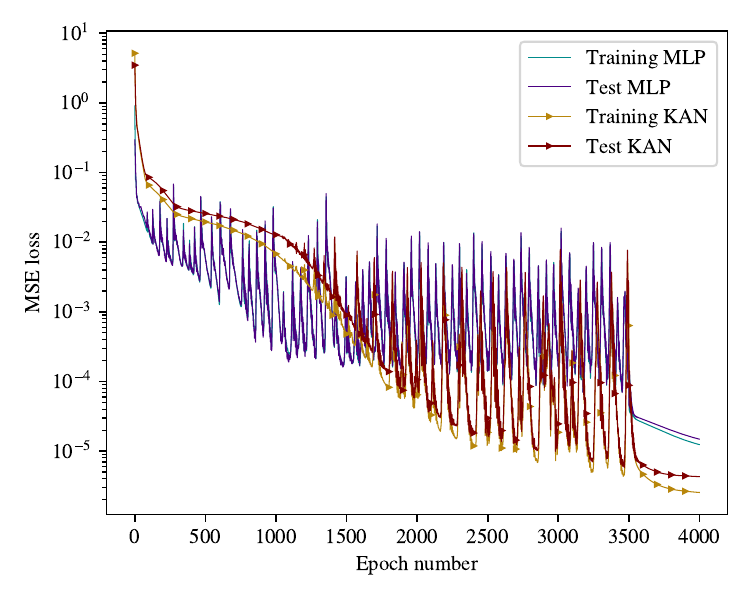}
            \caption{Laurent method.}
        \end{subfigure}
        \caption{Training and test loss curves for the experiment in \Cref{sec:test3}.}
        \label{fig:test3loss}
\end{figure}

 Finally, we display in \Cref{fig:test3loss} the loss curves during the training of the four neural networks. In this respect, PIHKANs exhibit slower convergence initially but improve more rapidly in the later stages.

\section{Conclusion and outlook}\label{sec:conclusions}
Some limitations of the original holomorphic learning method have been overcome through the use of effective strategies. First, it has been shown that Vekua operators can be integrated in the neural network representation to extend the application of the method to more general elliptic problems, including the Helmholtz equation. This could represent a turning point in the context of physics-based machine learning. On the other hand, little is known on Vekua operators for more general problems, both in terms of their explicit definition and whether the map is surjective. Another issue is represented by the numerical evaluation of the Vekua integral, which inevitably complicates the learning process and increases the computational cost. Future research may address a more efficient quadrature method or integral representation.

The Laurent-based method demonstrated to be suitable for learning solutions in multiply-connected domains, outperforming the domain decomposition method. In particular, convergence is more rapid and stable, while efficiency is comparable.  It should be noted again that the observed difference in training times reported in \Cref{tab:test3} is mainly due to the lower degree of optimization in the current code implementation.

In all tests, both types of holomorphic neural networks achieve great accuracy and efficiency, consistently outperforming traditional PINNs. In particular, PIHKANs exhibit better convergence compared to PIHNNs, even when employing smaller network architectures, although this sometimes leads to slightly longer training times. This was already remarked in previous studies and is presumably attributable to two different factors: first, the evaluation of KAN activation functions is inherently more complicated and requires multiple processing steps. In addition, current machine learning libraries may lack optimized implementations for these operations, particularly for GPU accelerated parallel execution. Although the first limitation appears intrinsic to the architecture, the second may be mitigated in future library releases, potentially narrowing the training time gap between PIHKANs and PIHNNs.

Overall, the holomorphic network approach has been significantly extended and enhanced through the strategies proposed in this work, making it a competitive alternative with respect to the current state of the art. While the finite element method remains more efficient for solving direct problems by taking a few seconds for the proposed benchmark tests, holomorphic networks offer a promising path toward narrowing the performance gap between physics-informed machine learning and traditional numerical solvers. Nonetheless, a key limitation persists in the current restriction to 2D problems, which may be overcome in future work by adopting more advanced mathematical representations.

\section*{CRediT authorship contribution statement}
Matteo Calafà: Writing – original draft, Writing - review \& editing, Software, Formal analysis, Conceptualization, Methodology, Validation, Investigation, Visualization. Tito Andriollo: Writing – review \& editing, Funding acquisition, Formal analysis. Allan P. Engsig-Karup: Writing – review \& editing, Conceptualization, Formal analysis. Cheol-Ho Jeong: Writing – review \& editing, Funding acquisition, Formal analysis.

\section*{Data availability}
Data will be made available on request.

\section*{Acknowledgments}
M.C.~was partly supported by the Aarhus University Research Foundation, Denmark, through the grant no.~AUFF-E-2023-9-44 “Strength of materials-informed neural networks”. In addition, this work partly contributes to the activities of the DTU Alliance PhD project of M.C.~“Scientific machine learning for 6 degree of freedom (DoF) augmented reality (AR) / mixed reality (MR) audio rendering”.

\section*{Declaration of competing interest}
The authors declare that they have no known competing financial interests or personal relationships that could have appeared to
influence the work reported in this paper.

\bibliographystyle{unsrtnat}
\bibliography{bibliography}
\end{document}